\documentclass[12pt]{article}
\usepackage{amsmath,amssymb}
\usepackage{graphics}
\usepackage[dvipdfm]{graphicx}
\usepackage{color}
\usepackage{url}
\makeatletter

\@addtoreset{equation}{section}
\makeatother

\makeatletter
\def\alt{\mathrel{\mathpalette\gl@align<}}
\def\agt{\mathrel{\mathpalette\gl@align>}}
\def\gl@align#1#2{\lower.6ex\vbox{\baselineskip\z@skip\lineskip\z@
\ialign{$\m@th#1\hfil##\hfil$\crcr#2\crcr\sim\crcr}}}
\makeatother

\begin{document}
\begin{titlepage}

\begin{center}

{\Large\bf 
Revisiting fermion mass and mixing fits in the minimal SUSY $SO(10)$ GUT
} 
\lineskip .75em
\vskip 1.5cm

\normalsize
{\large Takeshi Fukuyama},$^1$ 
 {\large Koji Ichikawa},$^2$
and
 {\large Yukihiro Mimura}$\,^3$

\vspace{1cm}

$^1${\it  Research Center for Nuclear Physics (RCNP), Osaka University, \\ 
Ibaraki, Osaka, 567-0047, Japan} \\

$^2${\it  University of Tokyo, \\ 
Kavli Institute for the Physics and Mathematics of the Universe (KIPMU) \\
5-1-5 Kashiwa-no-Ha, Kashiwa Shi, Chiba 277-8568, Japan} \\

$^3${\it Department of Physics, 
 National Taiwan University, 
 Taipei, 10617, Taiwan} \\

\vspace*{10mm}

{\bf Abstract}\\[5mm]
{\parbox{13cm}{\hspace{5mm}
%

Supersymmetric $SO(10)$ grand unified models with renormalizable Yukawa couplings involving only ${\bf 10}$ and
$\overline{\bf 126}$ Higgs fields have been shown to realize the fermion masses and mixings economically.
In previous works, the sum rule of the fermion mass matrices are given by inputting the quark matrices,
and the neutrino mixings are predicted in this framework. 
Now the three neutrino mixings have been measured, 
and in this paper, we give the sum rule by inputting the lepton mass matrices,
which makes clear certain features of the solution, especially if 
the vacuum expectation values of ${\bf 126}+ \overline{\bf126}$ ($v_R$) are large
and the right-handed neutrinos are heavy.
We perform the $\chi^2$ analyses to fit the fermion masses and mixings using the sum rule.
In previous works, the best fit appears at $v_R \sim 10^{13}$ GeV, and  
the fit at the large $v_R$ scale ($\sim 10^{16}$ GeV) has been less investigated.
Our expression of the sum rule has a benefit to understand the flavor structure in the large $v_R$ solution.
Using the fit results, we perform the calculation of the $\mu \to e\gamma$ process
and the electric dipole moment of electron, and the importance of $v_R$ dependence emerges in low energy phenomena.
We also show the prediction of the CP phase in the neutrino oscillations,
which can be tested in the near future.

\vspace*{7mm}

}}
\end{center}

\end{titlepage}


\baselineskip=16pt

\section{Introduction}

The Higgs discovery at the LHC \cite{Aad:2012tfa} opens a new era to understand the fermion masses and the mixings,
which are generated by the Yukawa interaction to the Higgs boson.
Indeed, most of the parameters in the standard model (SM) lie in the Yukawa coupling matrices.
As the remaining parameters in the flavor sector in the SM,
the neutrino 13-mixing angle has been measured 
by using the short baseline neutrino oscillations from the reactor neutrino \cite{An:2012eh},
and the CP phase in the three-flavor neutrino oscillations 
(the Pontecorvo-Maki-Nakagawa-Sakata (PMNS) phase) 
is expected to be measured accurately in the long baseline oscillations \cite{Wilking:2013vza}.
The preferable region has been already obtained by the global data analysis 
for the neutrino oscillations \cite{GonzalezGarcia:2012sz, Capozzi:2013csa,Capozzi:2016rtj}.
Do the accurate measurements of the parameters in the SM mean the end of the story?
The answer is of course No!
Even after the parameters are accurately measured, they do not fix the structure of the Yukawa matrices 
because of the existence of the unphysical flavor rotation.
The particle physicists claim that there must be new physics beyond the SM from various points of view.
The unphysical rotation in the SM can be physical in models beyond the SM.
In the era that the SM parameters in the flavor sector are accurately measured, 
we should, thus, study the structure of the Yukawa matrices
which can influence observables beyond the SM
to test the various flavor models.

The difference of the generation mixings between quark and lepton sectors
is one of the major issues to understand the structure in the Yukawa coupling matrices in the unified flavor picture.
The $SO(10)$ grand unified theory (GUT) provides a promising framework
to unify the quarks and leptons,
because the entire SM matter contents of each generation (including a right-handed neutrino)
can be unified in a single irreducible representation, $\bf 16$.
Particular attention has been paid to
the renormalizable minimal $SO(10)$ model, 
where two Higgs multiplets 
 $\{{\bf 10} \oplus {\bf \overline{126}}\}$ 
 are utilized for the Yukawa couplings with the matter representation \cite{Babu:1992ia}. 
The couplings to the {\bf 10} and $\overline{\bf 126}$ Higgs fields
can reproduce realistic charged fermion mass matrices using their phases thoroughly \cite{Matsuda:2000zp}.
Qualitatively, the smallness of the generation mixings in the quark sector can be explained by the left-right symmetry
which is a subgroup of the $SO(10)$ symmetry, while the neutrino mixings are not necessarily small
and large mixing angles for the solar and atmospheric neutrino oscillations can be obtained naturally in this framework.
Since the fermion mass matrices are given by the linear combination of two symmetric matrices 
(multiplied by doublet Higgs mixings),
one can obtain the sum rule of the fermion mass matrices.
Because the number of the physical parameters in the minimal model is restricted,
algebraic predictions of the SM parameters can be obtained using the sum rule.
Actually, the quantitative prediction of neutrino oscillation parameters
has been discussed by inputting the quark masses and mixings \cite{Fukuyama:2002ch,Bajc:2002iw}.
The model predicts that the neutrino 13-mixing angle is non-zero,
and the predicted value is consistent with the measured value.
We now repeat the question in the previous paragraph.
When the mixings and the CP phases in the neutrino oscillations are measured accurately,
does it mean the end of the study of the minimal SO(10) 
just by checking with the predictions after the number-crunching fits?

The supersymmetric (SUSY) theory is one of the most attractive candidates to construct a model
beyond the SM, and can explain the large hierarchy between the weak scale and the GUT scale.
In SUSY models, the structure of the Yukawa matrices can influence the low energy phenomena
if the SUSY particle masses are less than around a few TeV.
Even after the SM parameters are fitted (within the experimental errors), there can be physical degrees of 
freedom to influence the observables, though they also depends on the SUSY particle spectrum.
Therefore, it is important not only to fit the SM parameters, but also to understand the flavor structure
consistent with the experimental data in the format applicable to extended models.
From such a point of view,
in this paper, 
we revisit the fits of the fermion masses and mixings in the minimal SUSY $SO(10)$ model. 
We first describe the sum rule of the fermion mass matrices in terms of the charged lepton mass matrix 
$M_e$ and seesaw neutrino mass matrix ${\cal M}_\nu$ as inputs,
and express the quark mass matrices by them.
In fact, the quark masses and mixings can receive large radiative 
corrections at low energy threshold in the supersymmetric version of 
the model, and there can be large ambiguities. Whereas, the masses and mixings in 
the lepton sector are now more accurately measured and less ambiguous.
Therefore, it is essential to investigate the properties of the solution 
inputting the parameters in the lepton sector.
Using the expression by inputting the parameters in lepton sector, 
we perform a $\chi^2$ analysis to fit the quark masses and mixings.
Surely,
the $\chi^2$ result must be the same whichever input is chosen in the expression.
Our claim is that the algebraic expression can be very helpful to understand the structure of the 
Yukawa matrices which reproduces the experimental data.

The scale of ${\bf 126}+\overline{\bf 126}$ vacuum expectation value (vev), $v_R$,
is important for breaking the rank-5 $SO(10)$ symmetry down to the rank-4 SM gauge symmetry.
The scale is also important to know how the light neutrino mass is generated by the seesaw mechanism \cite{seesaw}.
In the $SO(10)$ model with $\overline{\bf 126}$ Higgs coupling,
not only the right-handed neutrino Majorana mass for type I seesaw, 
but also the $SU(2)_L$ triplet contribution
to the light neutrino mass is generated known as type II seesaw 
\cite{Schechter:1980gr}.
In GUT models, an intermediate scale often appears, and it can implement the proper size of the light neutrino masses.
In fact, 
the best fit in the minimal model is obtained when $v_R \sim 10^{13}$ GeV,
as was reported in Refs.\cite{Babu:2005ia,Bertolini:2006pe}.
However, if $v_R$ is about $10^{13}$ GeV in SUSY models,
the decomposed representations in ${\bf 126}+\overline{\bf 126}$ lie around the scale,
and they are harmful to the gauge coupling evolution
since gauge couplings may blow up before the unification.
Being motivated by this fact,
we feature the solutions for $v_R$ being around $10^{16}$ GeV.
In order to construct the solutions in which the type I seesaw part contributes the 
neutrino masses, the Yukawa coupling matrix $f$ for $\overline{\bf 126}$ Higgs coupling
should be nearly singular and the elements of $f^{-1}$ are enlarged.
Such a situation is not obvious if the quark masses and mixings are inputs.
Contrary to the situation, in our expression in terms of the parameters in the lepton sector,
it is easy to decode the structure of the singular matrix.
We show the results of the $\chi^2$ fit
by changing $v_R$, and also show the result of type I seesaw, where $SU(2)_L$ triplet contribution is negligible.
From the fit results,
we point out that future accurate measurements of the neutrino 23-mixing angle
and the PMNS phase play a decisive role to determine the Yukawa structure
in the solutions for $v_R \agt 10^{16}$ GeV.
This feature can be algebraically explained in our expression of the sum rule.

In the minimal $SO(10)$ model, 
the Dirac neutrino coupling and the Majorana neutrino mass matrix are predictable depending on the scale $v_R$.
As a consequence, by investigating the lepton flavor violation (LFV),
it is possible to find a footprint to know 
which scenario of the neutrino mass generation is chosen. 
We will perform the calculation of the branching fraction of $\mu \to e\gamma$ decay
and the electric dipole moment (EDM) of the electron,
using the prediction from the $\chi^2$ analysis in the model
and discuss the $v_R$ dependence of the predictions.

This paper is organized as follows:
In section 2, we introduce the sum rule of the fermion mass matrices in the $SO(10)$ model with the
minimal Yukawa interaction.
In section 3, 
the expression of quark mass matrices in terms of the mass and mixing parameters in the lepton sector is derived.
We examine the properties of the solutions expressed by the parameters in the lepton sector,
and study the $v_R$ dependence of the solutions.
We perform the numerical analyses to fit the fermion masses and mixings in section 4.
The importance of the predictions of the phase in neutrino oscillations is illustrated.
In section 5, we apply our fit results to the calculations of branching ratio of $\mu\to e\gamma$ process. 
We also mention the $v_R$ dependence of the results and study the importance to calculate the electron EDM.
We discuss the possible modification of the Yukawa couplings by the SUSY threshold corrections.
Section 6 is devoted to conclusions and discussions.
In the appendix, we express the square root matrix, which is necessary to express the solution in terms of the parameters of the lepton sector.

\section{Sum rule of the fermion mass matrices in the minimal $SO(10)$ model}

The SUSY GUT can provide the most promising framework
to incorporate the vast data systematically and consistently.
Among many candidates, $SO(10)$ \cite{Fritzsch} is the smallest simple gauge group 
 under which the entire SM matter contents of each generation are unified into a single anomaly-free irreducible representation, ${\bf 16}$. 
The ${\bf 16}$-dimensional spinor representation in $SO(10)$ includes the right-handed neutrino and no other exotic matter particles.

The minimal $SO(10)$ model\footnote{
In this paper, we define the minimality of the $SO(10)$ model on the Yukawa interaction:
The matter representations couple to only ${\bf 10}$ and $\overline{\bf 126}$ Higgs representations,
and the number of parameters in the Yukawa couplings is minimal (without a symmetry).
People often impose the minimality of the model even on the Higgs contents,
${\bf 10} + \overline{\bf 126} + {\bf 126} + {\bf 210}$,
 in which the number of parameters in the Higgs interactions is minimal (in addition to the minimality in the Yukawa interactions).
 In this case, however, the SUSY version of the model 
 suffers some delicate mismatches between all the fermion data fittings appeared in the 
 recently developed measurements and the self-consistency of the gauge coupling unification \cite{Bertolini:2006pe},
 and the framework suggests a non-SUSY model (or split SUSY) \cite{Bajc:2008dc,Bertolini:2009es}.
 Since we focus on the prediction on the Yukawa structure which can communicate with the low energy phenomena,
 we impose the minimality just on the Yukawa interactions.
 See section 6 for discussion.
} is defined as that in which the Yukawa interaction is minimal,
namely, only $H: {\bf 10}$ and $\bar\Delta: \overline{\bf 126}$ Higgs representations couple
to the fermions $\psi: {\bf 16}$ by renormalizable interaction:
\begin{equation}
W_Y= \frac12  {\bf h}_{ij} \psi_i \psi_j H + \frac12  {\bf f}_{ij} \psi_i \psi_j \bar \Delta.
\end{equation}
Due to the $SO(10)$ algebra, the coupling matrices are symmetric, ${\bf h}_{ij} = {\bf h}_{ji}$ and ${\bf f}_{ij} = {\bf f}_{ji}$.

The Higgs superpotential is investigated to see the detail pattern of symmetry breaking 
from $SO(10)$ down to the SM gauge group \cite{clark,Aulakh:2004hm}. 
In these analyses, the criterion of the renormalizability plays an essential role 
not only to reduce the number of parameters but also
to construct a model without the vevs of 
$B-L = \pm1$ direction by
 ${\bf 16}_H + \overline{\bf 16}_H$ representations to break $SO(10)$ symmetry.
Actually, the vev of $B-L=\pm2$ direction by ${\bf126}+\overline{\bf 126}$ fields reduces 
the rank of the gauge symmetry in the renormalizable model,
in which the $R$-parity is automatically conserved in the minimal SUSY standard model (MSSM) vacua.
The extension to include {\bf 120} Higgs representation
without breaking the renormalizability has been also considered \cite{120}. 
The $\overline{\bf 126}$ Higgs representation includes both $SU(2)_L$
and $SU(2)_R$ triplets, and the Yukawa interaction to $\overline{\bf126}$ can generate 
both left- and right-handed Majorana neutrino mass matrices.
Therefore, 
in the framework of the minimal $SO(10)$,
depending on the symmetry breaking vacua, 
the light neutrino mass matrix can be obtained \cite{Bajc:2002iw,Babu:2005ia} 
in both type I and II seesaw mechanism \cite{seesaw,Schechter:1980gr}.

The Yukawa coupling (after $SO(10)$ symmetry is broken down to the SM) is given as follows:
\begin{eqnarray}
&& Y_u = h + r_2 f, \nonumber \\
&& Y_d = r_1 (h + f), \nonumber \\
&& Y_e = r_1 (h - 3 f), \nonumber \\
&& Y_\nu = h- 3r_2 f,
\end{eqnarray}
where $r_1$ and $r_2$ depend on the Higgs mixing (doublet Higgs mixing in {\bf 10} and $\overline{\bf 126}$),
and 
$h$ and $f$ are original Yukawa matrices $\bf h$ and $\bf f$ multiplied by Higgs mixings\footnote{
The unitary matrices $U$ and $V$ are the diagonalizing matrices of the doublet Higgs matrix $M_{\rm doublet}$:
$U M_{\rm doublet} V^T$ is diagonal,
\begin{equation}
-{\cal L}_{\rm doublet} = (H_d^{10}, \bar\Delta_d, \Delta_d, \Phi_d) M_{\rm doublet} (H_u^{10}, \Delta_u, \bar\Delta_u, \Phi_u)^T.
\end{equation}
The lightest linear combinations of the doublets will be the MSSM Higgs doublets.
The detail can be found in Ref.\cite{Dutta:2005ni} (and \cite{Aulakh:2004hm} in different conventions).
}:
\begin{eqnarray}
h = V_{11} {\bf h}, \qquad
f = \frac{U_{12}}{\sqrt3 r_1} {\bf f}, 
\qquad
r_1 = \frac{U_{11}}{V_{11}},
\qquad
r_2 = r_1 \frac{V_{13}}{U_{12}}.
\end{eqnarray}

The charged fermion masses are obtained as
\begin{equation}
M_u = Y_u v_u, \qquad
M_d = Y_d v_d, \qquad
M_e = Y_e v_d, \qquad
M_\nu^D = Y_\nu v_u.
\end{equation}
where $v_u$ and $v_d$ are
the VEVs of up- and down-type Higgs fields.
We obtain the relation of the mass matrices as
\begin{eqnarray}
&& M_d = M_e + \frac{4}{1-r_2} F = r M_u + F, \\
&& r M_\nu^D = M_e + 3F,
\end{eqnarray}
where 
\begin{equation}
r = r_1 \frac{v_d}{v_u} \equiv r_1 \cot\beta,
\end{equation}
and the matrix $F$, which is proportional to the $\overline{\bf 126}$ Higgs coupling matrix, is
\begin{equation}
F = r(1-r_2) f v_u.
\end{equation}
Roughly, we obtain $r \sim m_b/m_t$.
Surely, these mass relations are realized at GUT scale, and the evolution via renormalization group equations (RGEs)
is considered to fit them with the low energy data of quark-lepton phenomena including the neutrino oscillations \cite{Fukuyama:2002ch}.

The right-handed Majorana neutrino mass matrix is obtained
as
\begin{equation}
M_R = \sqrt2 {\bf f} v_R,
\end{equation}
where $v_R$ is a VEV of $\overline{\bf 126}$.
Practically, we denote
\begin{equation}
M_R = c_R v_R f,
\end{equation}
where 
\begin{equation}
c_R = \sqrt6 \frac{r_1}{U_{12}} = \sqrt6 \frac{r_2}{V_{13}},
\end{equation}
 for the current notation.
Because $U_{12}$ and $V_{13}$ are components of the diagonalization unitary matrix for doublet Higgs fields,
$c_R$ has a minimal value.
The size of $c_R$ is related to the size of original ${\bf f}$ coupling,
which will be important to derive the GUT scale threshold correction
for flavor violation.
One can rewrite as
\begin{equation}
M_R = \frac{c_R v_R}{r(1-r_2) v_u} F.
\end{equation}

%

The seesaw neutrino mass matrix can be written as \cite{seesaw, Schechter:1980gr} 
\begin{equation}
{\cal M}_\nu = M_L - M_\nu^D M_R^{-1} (M_\nu^D)^T,
\end{equation}
where
$M_L$ is the left-handed neutrino Majorana mass which comes from $SU(2)_L$ triplet coupling \cite{Cheng:1980qt},
$\ell \ell \Delta_L$.
In $SO(10)$ model,
the $\overline{\bf 126}$ Higgs also includes the $SU(2)_L$ triplet $\Delta_L$
and the Yukawa coupling generates both $M_L$ and $M_R$.
Therefore, $M_L$ is also proportional to the coupling matrix $f$ and we denote 
\begin{equation}
M_L = c_L v_L f.
\end{equation}
In the $SO(10)$ model, 
if there is {\bf 210} or ${\bf 54}$ Higgs representation\footnote{
When ${\bf 54}$ is employed to generate the triplet contribution to the neutrino masses \cite{Goh:2004fy}, 
the $SO(10)$ breaking vacua is modified, and $r_2$ parameter can be independent to the vacua.
},
the VEV of $\Delta_L$, $\langle \Delta_L \rangle = v_L$, can be obtained as
$v_{\rm weak}^2/M_\Delta$, where $M_\Delta$ is the mass of the $SU(2)_L$ triplet.

For convenience, we reparameterize $M_R$ and $M_L$ as 
\begin{eqnarray}
M_R &=& \frac{1}{r^2 R} F \times 10^{10}, \\
M_L &=& \delta F \times 10^{-10}, \label{M_L}
\end{eqnarray}
where
\begin{equation}
R \equiv \frac{1-r_2}{r} \frac{v_u}{c_Rv_R} \times 10^{10}, \qquad \delta \equiv \frac{c_L v_L}{r(1-r_2) v_u} \times 10^{10}.
\end{equation}
The factor $10^{10}$ is attached for a numerical convenience. 
We will treat $R$ as a parameter to specify the scale of $v_R$ implicitly.
For example, if $v_R \sim 10^{13}$ GeV, one obtains $R \sim O(1)$ roughly.
For $v_R \sim 10^{16}$ GeV, one finds $R \sim O(10^{-3})$.

We remark on the naive statement on the scale $v_R$ to fit the fermion masses and mixings,
which can be found in the previous works.
%
%
One may think that the 0.05 eV neutrino mass can be easily obtained
even if $v_R \sim 10^{16}$ GeV
because the scale of $M_R = c_R v_R f$ can be naturally $O(10^{14})$ GeV
for the component of $f$ to be $O(10^{-2})$.
However, such a naive thought is true only if the atmospheric mixing is small.
In fact, in order to reproduce a large atmospheric mixing,
not only $({\cal M}_\nu)_{33}$ but also $({\cal M}_\nu)_{23}$ should be $O(0.01)$ eV.
In the $M_e$-diagonal basis,
one obtains $(M_e F^{-1} M_e)_{23} \ll (M_e F^{-1} M_e)_{33}$,
and thus, to obtain a large 23-mixing, one needs a cancellation in (3,3) element of ${\cal M}_\nu$,
and the naive thought above does not work.
To obtain the large atmospheric mixing,
$(M_e f^{-1} M_e)_{23}/(c_R v_R)$ has to be $O(0.01)$ eV.
Because the size of $f$ is $O(0.01)$ due to the fitting of quark masses and mixings,
the naive size of the $v_R$ is $O(10^{13})-O(10^{14})$ GeV,
which corresponds to $R\sim O(1)$.
One can find a solution for $v_R \sim O(10^{16})$ GeV (i.e. $R \sim 10^{-3}$),
only if the matrix $f$ is nearly singular and $f^{-1}$ is enhanced (compared to its naive size from 
each component).

We note that the solution for the nearly singular $f$ matrix is also found in Ref.\cite{Bertolini:2006pe} (they call the solution
as ``Mixed$^{\prime}$'').
However, if one fits the charged fermion masses, there is no reason that 
$f$ is close to a singular matrix,
and it is not easy to study the property of the solution.
We describe the neutrino mass and mixing parameters to be inputs,
and construct a solution where the situation of small $\det f$ can be seen explicitly.
The solution of nearly singular $M_R$ matrix is applicable
even to the non-minimal models.
We believe that the description of nearly singular $M_R$ is useful 
to study how the light neutrino mass scale is obtained in seesaw models.

\section{Property of the solution by inputting the lepton parameters}

We have described the general setup in the minimal $SO(10)$ model.
In the previous works, the fitting of the fermion masses and mixings 
has been performed by inputting the quark masses and mixings, and the neutrino mixings 
are predicted in the framework.
At present, the lepton parameters are more accurate rather than the quark ones,
and besides, a large threshold correction are expected in the quark sector in SUSY models.
In that sense, it is better to perform the fitting by inputting the parameters in lepton sector,
and outputting the quark parameter.
The formulation is presented in not only such a practical purpose, but also 
to make clear the property of the solution with $v_R \sim 10^{16}$ GeV, which is main concern in this paper.

Using the relation, $rM_\nu^D = M_e + 3F$,
we obtain
\begin{eqnarray}
{\cal M}_\nu \times 10^{10} &=& \left( \delta F - R (M_e+ 3F)F^{-1} (M_e + 3F) \right) \nonumber\\
&=&\left( (\delta-9R) F - 6R M_e - R M_e F^{-1} M_e \right).
\end{eqnarray}
This equation can be rewritten as a quadratic equation in terms of $M_e^{-1/2} F M_e^{-1/2}$, 
and we obtain
\begin{equation}
F = 
\frac{1}{18 R - 2\delta} M_e^{1/2} \left(
 K - 6 R \, {\bf 1} + \sqrt{ K^2 - 12 RK + 4 \delta R \,{\bf1}}
 \right) M_e^{1/2},
 \label{F-matrix}
\end{equation}
where
\begin{equation}
K \equiv - M_e^{-1/2} \hat{\cal M}_\nu M_e^{-1/2}, \qquad \hat{\cal M}_\nu \equiv {\cal M}_\nu \times 10^{10}.
\end{equation}
The square root matrix is defined in the Appendix.
Using the expression of $F$ matrix,
the quark mass matrices, $M_u$ and $M_d$,
can be given as a function of charged lepton mass matrix $M_e$ and the light neutrino mass matrix ${\cal M}_\nu$.

Now, 
by using the formula given in Appendix, let us express the matrix 
\begin{equation}
K- 6R {\bf 1} + \sqrt{(K-6R{\bf1} )^2 + D{\bf1}},
\end{equation}
where $D \equiv 4\delta R - 36 R^2$.
Surely because $K$ and $(K-6R {\bf1})^2 +D {\bf1}$ can be diagonalized
simultaneously,
we can easily obtain
\begin{equation}
K- 6R {\bf 1} + \sqrt{(K-6R{\bf1} )^2 + D{\bf1}}
= \sum_i \left(\lambda_i -6R+ s_i \sqrt{(\lambda_i - 6R)^2 +D}\right) \Lambda_i ,
\end{equation}
where $\lambda_i$ $(i = 1,2,3)$ are eigenvalues of matrix $K$,
$s_i$ are signs for the square roots, and $\Lambda_i$ are given in Appendix.
By the definition of the matrix $K$, $K = - M_e^{-1/2} \hat{\cal M}_\nu M_e^{-1/2}$ 
(where $\hat {\cal M}_\nu = {\cal M}_\nu \times 10^{10})$,
we obtain
\begin{equation}
M_e^{1/2} \Lambda_1 M_e^{1/2}
= 
\frac{1}{(\lambda_1 - \lambda_2)(\lambda_1 - \lambda_3)}
\left(
 \hat{\cal M}_\nu M_e^{-1} \hat{\cal M}_\nu  + (\lambda_2 + \lambda_3) \hat{\cal M}_\nu + \lambda_2 \lambda_3 M_e
\right),
\end{equation}
and similarly for $\Lambda_{2,3}$.

In the limit of $R\to 0$ (i.e. $v_R \to \infty$),
we obtain
\begin{equation}
F \to - \frac{1}{2\delta} M_e^{1/2} \left(K + \sqrt{K^2} \right) M_e^{1/2}.
\end{equation}
So, let us evaluate
$K + \sqrt{K^2}$.
One can write down
  \begin{equation}
K + \sqrt{K^2} = V 
\left(
  \begin{array}{ccc}
   (1+s_1)\lambda_1 & & \\
   & (1+s_2)\lambda_2 & \\
   & & (1+s_3)\lambda_3 \\
  \end{array}
 \right)
 V^{-1}
 = \sum_i (1+s_i) \lambda_i \Lambda_i.
\end{equation}
For example, suppose 
$s_1 = -1$ and $s_{2,3}=1$
(because the square root of a complex number has two branches, 
we choose a sign convention $\sqrt{\lambda^2} = \lambda$ to define $s_i$),
one finds \begin{equation}
K + \sqrt{K^2} = 2\sum_{i = 2,3} \lambda_i \Lambda_i
= 2K - 2 \lambda_1 \Lambda_1.
\end{equation}
As a result, we obtain under the choice of $s_i$ and in the limit of $R \to 0$,
\begin{eqnarray}
F &=& \frac1{\delta} \left( 
\hat{\cal M}_\nu
+ \lambda_1 M_e^{1/2} \Lambda_1 M_e^{1/2}\right) \\
&=&
\frac1{\delta}\frac{1}{(\lambda_1 - \lambda_2)(\lambda_1 - \lambda_3)}
\left(
\lambda_1 \hat{\cal M}_\nu M_e^{-1} \hat{\cal M}_\nu  + (\lambda_1^2+\lambda_2\lambda_3) \hat{\cal M}_\nu + \lambda_1 \lambda_2 \lambda_3 M_e
\right).
\end{eqnarray}
In the choice of $s_1 = -1$, $s_2 = s_3= 1$,
 it can be written as
\begin{equation}
K + \sqrt{K^2} = 2V 
\left(
  \begin{array}{ccc}
   0 & & \\
   & \lambda_2 & \\
   & & \lambda_3 \\
  \end{array}
 \right)
 V^{-1}.
\end{equation}
Therefore, 
 ${\rm rank} (K+\sqrt{K^2})=2$  in this case. 
 As a consequence, one of the right-handed neutrinos is very light 
 compared to the other two because the right-handed Majorana mass matrix
 is proportional to $F$.
 Needless to say, $R$ is finite (though it is small), 
 and therefore, the lightest right-handed neutrino is not massless.

For a choice of $s_1= s_2= s_3= 1$,
one obtains
\begin{equation}
K + \sqrt{K^2} = 2K,
\end{equation}
and thus, in the limit of $R\to 0$,
\begin{equation}
F \to \frac{1}{\delta} \hat{\cal M}_\nu.
\end{equation}
Therefore, this choice corresponds to the case where the right-handed neutrinos are decoupled,
and the triplet contribution is dominant (namely ${\cal M}_\nu = M_L$).
It is known that the fits of fermion masses and mixings are not good in the triplet-dominant case \cite{Bertolini:2006pe}.
The difference compared to the previous case comes from the contribution from $\lambda_1 M_e^{1/2} \Lambda_1 M_e^{1/2}$,
which makes the fits better.
In our description of the solution,
the triplet-dominant case is contained in the expression just by a choice of the signs of the square root matrix.
In fact, as we execute $\chi^2$ analysis in the following sections, 
the triplet-dominant case does not appear in the fits by $\chi^2$ minimum, because it gives a large $\chi^2$ value.
Therefore, we call the combination of the triplet and type I seesaw contributions just as type II solution,
which is intrinsic in the $SO(10)$ model.
In the Ref.\cite{Bertolini:2006pe}, they name the solutions as ``Mixed'' and ``Mixed$^\prime$'',
but those two solutions are continuously connected as a function of the parameter $R$ in our description.

We note that the dependence of $R$ is small
in the type II solution for $s_1= -1$, $s_2= s_3=1$ for small $R$ (namely, large $v_R$)
as one can find from the $R\to 0$ limit.
In type I seesaw ($\delta=0$),
the $F$ matrix in the limit of $R \to 0$ can be written as
\begin{equation}
F \to \frac{1}{18R} M_e^{1/2} \left(K + \sqrt{K^2} \right) M_e^{1/2},
\end{equation}
and it surely depends on $R$.
For the solution $s_1 = s_2= s_3 =1$,
one obtains
\begin{equation}
F \to -\frac{1}{9R} \hat{\cal M}_\nu.
\end{equation}
For a smaller value of $R$, the component of $F$ becomes large and it is expected that there is no solution.
For a choice of $s_1 = s_2= s_3 = -1$,
one obtains\footnote{
Because of the algebraic equation $(K - 6 R \, {\bf 1} + \sqrt{ K^2 - 12 RK + 4 \delta R \,{\bf1}}) (K - 6 R \, {\bf 1} - \sqrt{ K^2 - 12 RK + 4 \delta R \,{\bf1}}) 
= 2R(18R - 2\delta) {\bf1}$,
one obtains
\begin{eqnarray}
F = \frac{1}{18R-2\delta} M_e^{1/2} (K - 6 R \, {\bf 1} - \sqrt{ K^2 - 12 RK + 4 \delta R \,{\bf1}}) M_e^{1/2} \nonumber\\
= 2R M_e^{1/2} (K-6R\,{\bf1} +\sqrt{K^2 - 12 RK + 4 \delta R \,{\bf1}})^{-1} M_e^{1/2}.
\end{eqnarray}
}
\begin{equation}
F
\simeq R M_e^{1/2} K^{-1} M_e^{1/2}
= - R M_e {\cal M}_\nu^{-1} M_e.
\end{equation}
For $R\to 0$, $F$ becomes small in this choice and which means 
there is up-down symmetry ($M_d \propto M_u$) in the limit,
and there is no solution.
Therefore, there may no solution for $v_R > O(10^{16})$ GeV in type I.

\section{Numerical verification}

In the previous section, 
we have studied the property of the solution for the case of a large $v_R$ ($R\to 0$).
Now, we verify the property by performing the numerical fitting of the fermion masses and mixings.
In our formulation, the lepton parameters are inputs and the quark parameters are outputs.
Of course, in order to perform the numerical fit in the practical way, we use the observed quark masses and mixings
to fit by $\chi^2$ analyses.
Therefore, the fit results have to be the same as the one given in the previous works (up to the detail updated experimental data) 
mathematically,
and the formulation itself does not provide a better fitting for sure.
As described, the purpose to use the expression by inputting the lepton parameters is to make clear the property of the solution
depending on the $v_R$, which is important to apply the model to the low energy physics.
In fact, the fitting in the minimal $SO(10)$ model needs complicate tuning in the parameters (depending on the parameterization)
especially in the case of a large $v_R$,
and one may feel that a better fit can exist by using the tuning of the parameters in quark sector.
However, in our expression, the solution provides a simple behavior for the large $v_R$ case,
and we can understand how the fit behaves.

In this section, we demonstrate the numerical fit of the fermion masses and mixings,
and we check the consistency to the previous works in the literatures.
Compared to era when the previous works are done in the literatures, the neutrino oscillation experiments 
enters to a new stage to measure the oscillation parameters:
not only the measurement of the 13-mixing angle, but also the expectation of the precision measurements of 23-mixing angle
and PMNS CP phase.
We first describe our parameterization which is suited in our scheme,
and next explain the method of our analyses including a technical detail. 
As a purpose to revisit the analyses,
we show our fit results making clear the importance of the precise measurement of 23-mixing angle.
Although the fit of the PMNS phase is constrained mildly, the correlation between the PMNS phase and 23-mixing angle 
can provide an important implication to distinguish models and the flavor structure in the different essential parameter regions.

\subsection{Parametrization}

The relation is summarized as
\begin{eqnarray}
r M_u &=& M_e + \frac{3+r_2}{1-r_2} F, \\
M_d &=& r M_u + F, \\
F &=& \frac{1}{18R - 2\delta} M_e^{1/2} \left(K- 6R {\bf1} + \sqrt{K^2 - 12 RK + 4\delta R {\bf1}}\right) M_e^{1/2},\\
K &=& - M_e^{-1/2} \hat{\cal M}_\nu M_e^{-1/2}.
\end{eqnarray}
We work on the basis where the charged lepton mass matrix $M_e$ is diagonal:
\begin{equation}
M_e = {\rm diag}.\, (m_e e^{i\alpha_e}, m_\mu e^{i\alpha_\mu}, m_\tau e^{i\alpha_\tau}),
\end{equation}
The light neutrino mass matrix is parametrized as
\begin{equation}
{\cal M}_\nu = \bar U {\rm diag.}\, (m_1 e^{i\alpha_1}, m_2 e^{i\alpha_2}, m_3 e^{i\alpha_3}) \bar U^T,
\end{equation}
and the unitary matrix $\bar U$ is the PMNS neutrino matrix, and we use the usual convention by particle data group (PDG).
In the matrix $\bar U$, there are three neutrino mixing angles $\theta_{12}, \theta_{23}, \theta_{13}$ and a $CP$ phase $\delta_{\rm PMNS}$
in neutrino oscillation.
As a convention, the mass parameters, $m_e, m_\mu, m_\tau$, $m_1,m_2,m_3$ are given as real and positive values.

As one can find easily, one of the phases $\alpha_e,\alpha_\mu,\alpha_\tau, \alpha_1,\alpha_2,\alpha_3$ can be made to be zero
without loss of generality,
since the relation is covariant under
\begin{equation}
M_{u,d,e} \to e^{i\alpha} M_{u,d,e}, \quad {\cal M}_\nu \to e^{i\alpha} {\cal M}_\nu.
\end{equation}
We choose $\alpha_1 = 0$ as a convention.
The relation is also covariant under the rephasing:
\begin{equation}
R \to e^{i\alpha} R, \quad \delta \to e^{i\alpha} \delta, \quad M_{u,d,e} \to e^{-i\alpha} M_{u,d,e},
\end{equation}
and thus we can choose $R$ to be real (and positive) without loss of generality.
We can also define the parameter $r$ to be real.
Therefore, the parameters in this parametrization are the following:
\begin{eqnarray}
m_e, m_\mu, m_\tau, m_1, m_2, m_3, \nonumber\\
\alpha_e, \alpha_\mu, \alpha_\tau, \alpha_2, \alpha_3, \nonumber\\
\theta_{12}, \theta_{23}, \theta_{13}, \delta_{\rm PMNS}, \\
r_2 (\mbox{complex}), r, \nonumber\\
\delta (\mbox{complex}), R,\nonumber
\end{eqnarray}
and there are 21 degrees of freedom in total.

The input parameters in the lepton sector are
\begin{equation}
m_e, m_\mu, m_\tau, \theta_{12}, \theta_{13}, \theta_{23},  \Delta m_{\rm sol}^2, \Delta m_{\rm atm}^2, \delta_{\rm PMNS},
\end{equation}
and the PMNS phase $\delta_{\rm PMNS}$ will be treated as output in our analysis.
The parameters in the quark sector are
\begin{equation}
m_u, m_c, m_t, m_d, m_s, m_b, V_{us}, V_{cb}, V_{ub}, \delta_{\rm KM},
\end{equation}
which we will fit by using the freedom in the model.
Totally, there are 19 degrees for the observables\footnote{
Depending on the size of absolute neutrino mass, the neutrino Majorana phases $\alpha_2$ and $\alpha_3$
and the neutrino mass $m_1$ can be the observables in the future by neutrinoless double beta decay
and the radiative emission of neutrino pairs from excited atoms \cite{Yoshimura:2013wva},
but we do not count them as observables in this paper. 
}.

Just from the degrees of freedom, 
one can fit all the 19 parameters using the full degrees of freedom in the model, in principle.
However, the parameters we need to fit is hierarchical 
and some of the degrees are phases, and consequently,
the fits do not necessarily done contrary to the simple number counting expectation.
Actually,
due to the hierarchy $m_u/m_t \ll m_e/m_\tau \ll m_d/m_b$,
the complex freedom $r_2$ should be almost consumed to fit the up-quark mass $m_u$
(roughly, the solution is obeyed by a cubic equation of $r_2$, $\det M_u \to 0$). 
Thus, roughly speaking, to fit $m_u$, two degrees of freedom is consumed.
As we have explained, 
the fit does not depend on $R$ for small $R$ very much,
and then the phase of $\delta$ is not an active freedom due to the rephasing covariance described above.
Therefore, the active number of parameters to fit the quark masses and mixings are reduced.
As a result, for the case of small $R$, down-quark mass $m_d$ is not fully fit as we will see later,
though $R\sim O(1)$ ($v_R \sim 10^{13}$ GeV) solution in type II can reproduce $m_d$, as it was known in literature \cite{Bertolini:2006pe}.

\subsection{Method of $\chi^2$ fit}

Before going to the detailed description of the fit process,
we explain the equations to solve the quark masses which we can find quickly
from the expression,
\begin{equation}
r M_u = M_e + \frac{3+r_2}{1-r_2} F =
M_e^{1/2} \left( {\bf 1} + \frac{3+r_2}{1-r_2} \frac{K-6R{\bf1} +\sqrt{(K-6R{\bf1})^2+D{\bf1}}}{18R-2\delta}
\right) M_e^{1/2}.
\end{equation}
As we have already noted,
due to the hierarchy $m_u/m_t \ll m_e/m_\tau$,
the matrix in the bracket in the above equation has to be nearly singular.
Therefore,
\begin{equation}
\frac{1-r_2}{3+r_2} + \frac{\lambda_i - 6R + s_i \sqrt{(\lambda_i-6R)^2+D}}{18R-2\delta} \to 0
\end{equation}
for one of $i=1,2,3$, where $\lambda_i$ are eigenvalues of the matrix $K$
as was given previously.
By perturbation from the above relation, the up-quark mass $m_u$ can be easily fit.
One can also easily solve the equation by $r$:
\begin{equation}
 r^2 (m_u^2+m_c^2+m_t^2) = r^2 \,{\rm Tr}\, M_u M_u^\dagger = \,{\rm Tr}\,\left(M_e + \frac{3+r_2}{1-r_2}F\right)\left(M_e + \frac{3+r_2}{1-r_2}F\right)^\dagger,
\end{equation}
and thus, roughly speaking, the top quark mass $m_t$ can be easily fit by using $r$.
As we have explained, we will try to fit the masses and mixings in quark sector
by inputting the masses and mixings in lepton sector.
The parameters to fit are now
\begin{equation}
m_c, m_d, m_s, m_b, V_{us}, V_{cb}, V_{ub}, \delta_{\rm KM}.
\end{equation}

The $\chi^2$ function is defined as 
\begin{equation}
 \chi^2 = \sum_i  \frac{(\chi_i - \hat\chi_i)^2}{\hat \sigma_i^2},
\end{equation}
where $\hat\chi_i$ and $\hat\sigma_i$ are the experimental measurements of the parameters and
their standard deviations of errors.

\begin{table}[t]
\begin{minipage}[t]{.45\textwidth}
\begin{tabular}{|c|c|c|}
\hline
\multicolumn{3}{|c|}{ Fixed Values } \\
\hline
$m_{u}$       & $ 3.961  \times 10^{-4} ~{\rm GeV} $ &  \cite{Bora:2012tx} \\
$m_{t}$       & $ 71.0883               ~{\rm GeV} $ &  \cite{Bora:2012tx} \\
$m_{e}$       & $ 3.585  \times 10^{-4} ~{\rm GeV} $ &  \cite{Bora:2012tx} \\
$m_{\mu}$     & $ 7.5639 \times 10^{-2} ~{\rm GeV} $ &  \cite{Bora:2012tx} \\
$m_{\tau}$    & $ 1.3146                ~{\rm GeV} $ &  \cite{Bora:2012tx} \\
$\Delta m_{\rm sol}^2$    & $ 7.54 \times 10^{-5} ~{\rm eV}^2 $   &  \cite{Capozzi:2013csa} \\
$\Delta m_{\rm atm}^2$    & $ 2.4  \times 10^{-3} ~{\rm eV}^2 $   &  \cite{Capozzi:2013csa} \\
$\theta_{12}$    &    0.583           &  \cite{Capozzi:2013csa} \\
$\theta_{23}$    &    0.710, $\pi/4$  &  \cite{Capozzi:2013csa} \\
$\theta_{13}$    &    0.156           &  \cite{Capozzi:2013csa} \\
\hline
\end{tabular}
\end{minipage}
\begin{minipage}[t]{.45\textwidth}
\begin{tabular}{|c|c|c|}
\hline
\multicolumn{3}{|c|}{ Parameters to fit } \\
\hline
$m_c$    & $ 0.1930  \pm 0.025              ~{\rm GeV} $   &    \cite{Bora:2012tx}    \\
$m_d$    & $ (9.316   \pm 3.8)   \times 10^{-4} ~{\rm GeV} $    &    \cite{Bora:2012tx}    \\
$m_s$    & $ (1.76702 \pm 0.5)   \times 10^{-2} ~{\rm GeV} $    &   \cite{Bora:2012tx}    \\
$m_b$    & $ (0.9898  \pm 0.03)                 ~{\rm GeV} $    &    \cite{Bora:2012tx}    \\
$     V_{us}$    &    $ 0.224 \pm 0.002 $ &    \cite{Agashe:2014kda}    \\
$     V_{cb}$    &    $ (3.7 \pm 0.13)  \times 10^{-2} $    &  \cite{Agashe:2014kda} \\
$     V_{ub}$    &    $ (3.7 \pm 0.45)  \times 10^{-3} $    &  \cite{Agashe:2014kda} \\
$\delta_{\mathrm{KM}}$    &   $ 1.18 \pm  0.2 $    &  \cite{Agashe:2014kda}  \\
\hline
\end{tabular}
\end{minipage}
\caption{\sl \small 
The reference parameters used in the fit.
}
\label{table:parameters}
\end{table}

The values and uncertainties used in the fit
are summarized in Table \ref{table:parameters} where 
we use the value of the quark mass in Table III of Ref.\cite{Bora:2012tx} (MSSM, $\tan \beta = 10$)
and calculate the quark mixing angle and KM phases at GUT scale
from PDG data \cite{Agashe:2014kda} by $\overline{{\rm DR}}$ scheme.

For the minimization, we use the Metropolis-Hastings (MH) algorithm~\cite{Metropolis:1953am} of the Markov Chain Monte Carlo (MCMC) method.
Since we expect that there exist many local minima, 
we perform the replica exchange MCMC (REMC) sampling~\cite{REMC} where several copies of the Markov chain 
with various temperature $T_{i}$ are simultaneously simulated.
Each chain performs MH scheme under a likelihood function defined by $\exp[- \chi^2/(2 T_{i})]$.
Every after performing some MC steps, the state of the chains are swapped by comparing the values of the likelihood.
For ordinary MCMC, the state of the chain is often trapped at a local minimum and 
 costs a large amount of time when we search a likelihood function with many local minima.
For REMC, on the other hand, 
a replica with high temperature ($> 1$)
has a flatter distribution function 
and the sampling point can easily move on to another local minimum.
Thanks to the swapping algorithm, this transition can propagate to low-temperature replicas 
and therefore we can effectively sweep all the local minima.

We should note, however, that it is still challenging to find the absolute global minimum 
of the likelihood function with many local deep minima. 
We have made much effort to find the minimum 
including changing the initial parameter set,
tuning the replica temperatures as well as the width of the parameter jump in MCMC sampling.
Nevertheless, there is always a possibility that the likelihood has another better minimum
 due to our limited hardware and calculation time.

The fits are done assuming that $M_{L}$ is negligible (type I: $\delta = 0$) or non-negligible (type II: $ \delta \neq 0$)\footnote{
The parameter $\delta$ is given in Eq.(\ref{M_L}),
and the classification of the $\chi^2$ minimal solution is described in section 3. 
}
under two values of $\theta_{23}~(0.710,~\pi/4)$.
In the MCMC, we fix the scale parameter $R$ to precisely investigate the $\chi^2$ behavior against the scale parameter, and 
therefore, the $\chi^2$ function has seven (nine) free parameters for type I (type II).\footnote
{
We also search the global best fit and the fit at $R = 0.001$ allowing the neutrino oscillation parameters 
($  \theta_{12}, \theta_{13}, \theta_{23} $) and neutrino mass difference 
($ \Delta m_{\rm sol}^2, \Delta m_{\rm atm}^2 $) free within a range of $3\sigma$.
We find that most of the parameters except for $\theta_{23}$ converge around their central values and
we decide to fix these reference values.

For $\theta_{23}$, on the other hand, although the $\theta_{23}$ deviation does not significantly change the $\chi^2$ values,
its uncertainty is still large.
Moreover, the $\theta_{23}$ value has a correlation with $\delta_{\rm{PMNS}}$ at the $\chi^2$ minimum 
as one can see in section 4.4. 
Therefore, in our fit, we adopt two values of $\theta_{23}$ as the reference value, 
where one is the global fit of the three neutrino oscillation parameters \cite{Capozzi:2013csa} ($\theta_{23} = 0.71$) and 
the other is the maximal mixing ($\theta_{23} = \pi/4$).
}
We adopt uniform prior for the phases
$( \alpha_{e}, \alpha_{\mu}, \alpha_{\tau}, \alpha_{2}, \alpha_{3}, \delta_{{\rm PMNS}}, {\rm Arg}(\delta))$
from $-\pi$ to $\pi$ and the magnitude of the seesaw parameter $|\delta| \leq 500$.
Log flat prior is applied to $m_1$  over the range of $10 \leq -\log_{10} (m_{1} / {\rm GeV} ) \leq 16 $.

\subsection{Fit results}

\begin{figure}[t]
\includegraphics[width=0.48\textwidth]{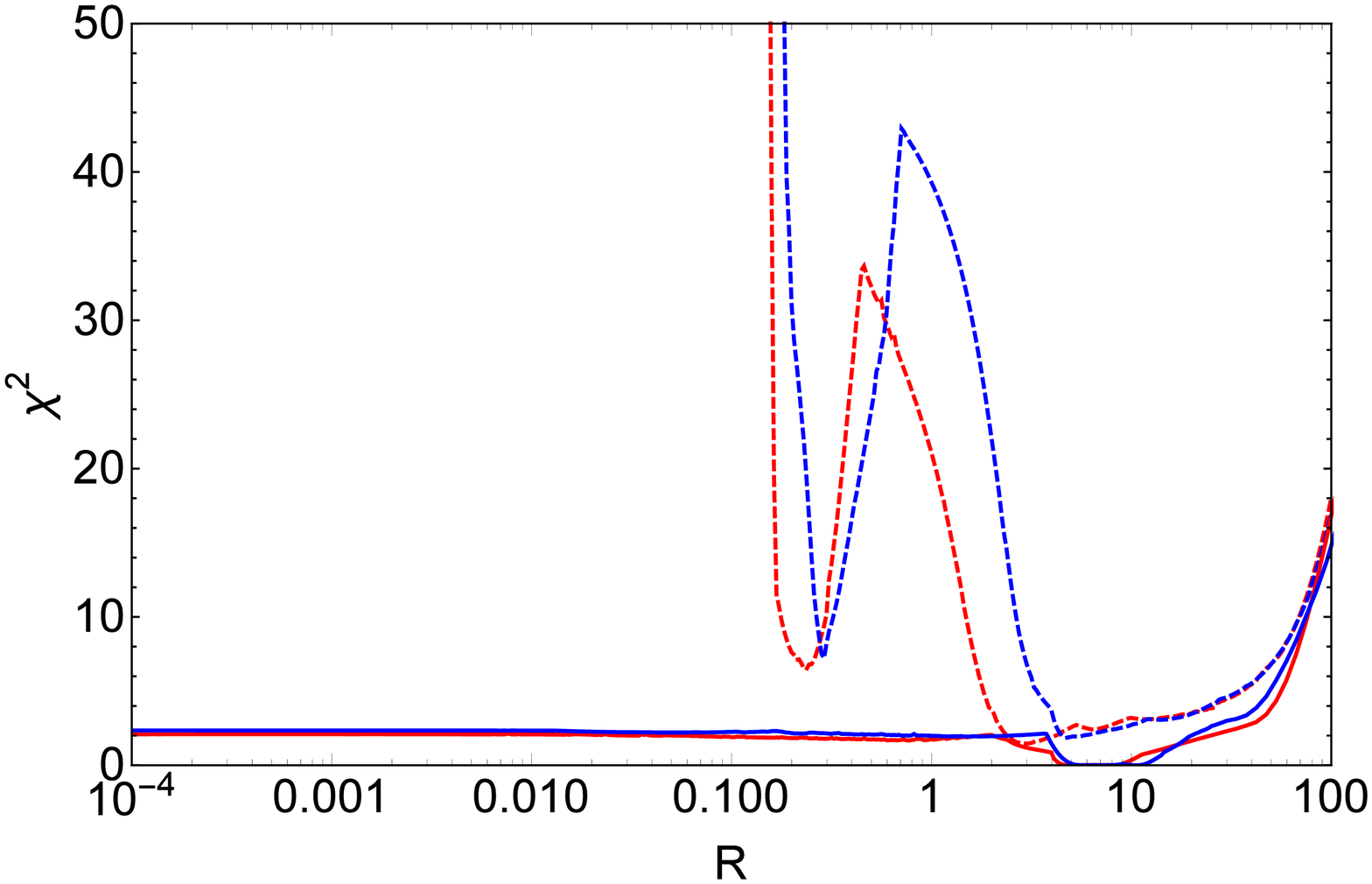}
\includegraphics[width=0.48\textwidth]{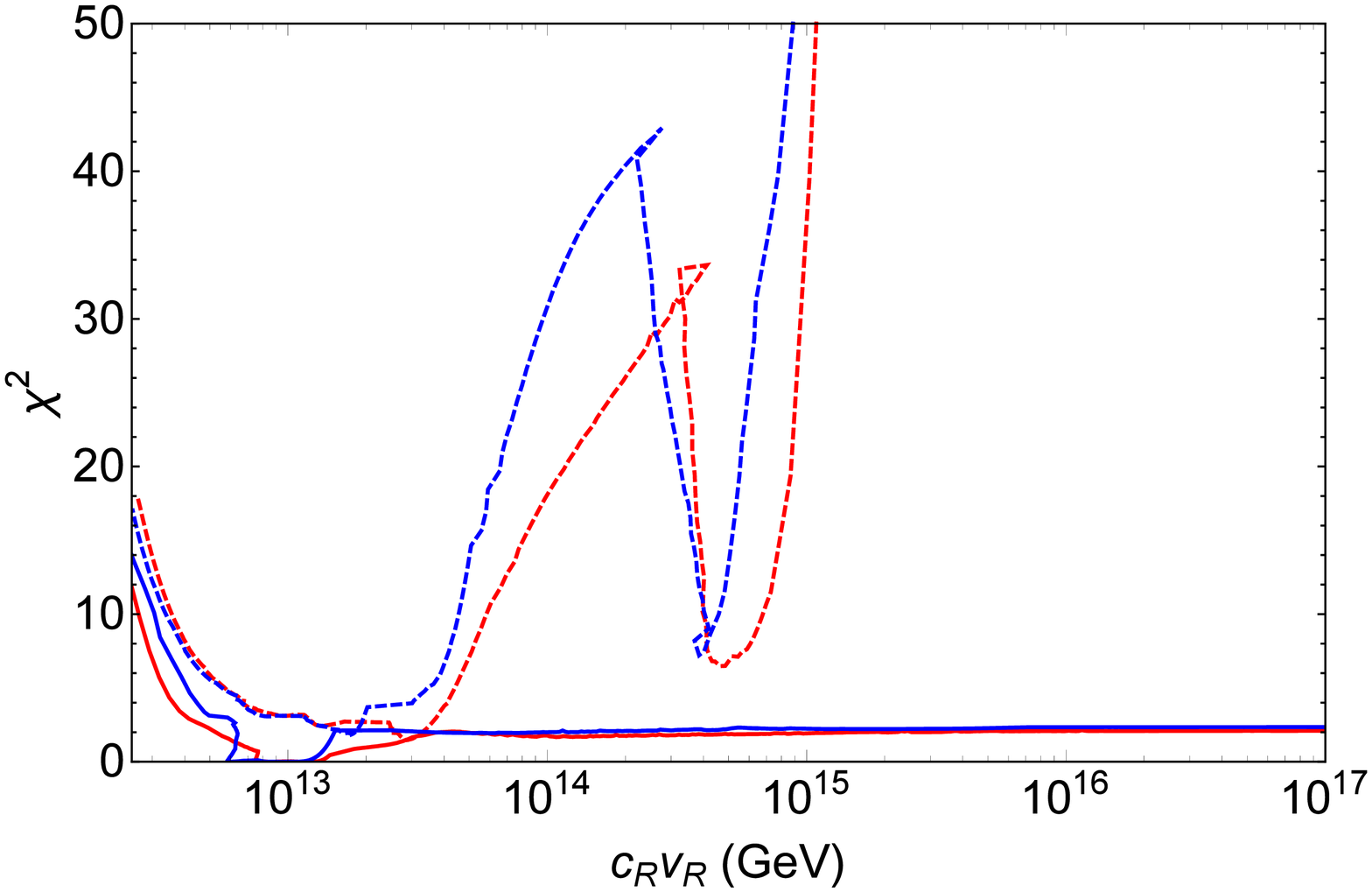}
\caption{\sl \small 
{\bf Left panel:} $R$ dependence of $\chi^2$ for 
type I of $\theta_{23} = 0.710$ (red dashed line), $\pi/4$ (blue dashed line)
and type II of $\theta_{23} = 0.710$ (red solid line), $\pi/4$ (blue solid line).
{\bf Right panel:} The same figure as the left one but 
we transform $R$ to the energy scale $c_Rv_R$.
}
\label{fig:bestfits}
\end{figure}

Fig.~\ref{fig:bestfits} shows $R$ (and $c_Rv_R$) dependence of $\chi^2$ values.
For all cases, the global best fit is found at around $v_R \sim O(10^{13})~{\rm GeV}$
where the $\chi^2$ value is 1.5\,--\,2 for type I and $\leq 0.001$ for type II.
The $\chi^2$ value of type I mainly stems from down quark mass ($\sim 0.44~{\rm MeV}$)
which is smaller than the experimental value by $\sim 1.2$\,--\,$1.3\sigma$.
For type I, another local minimum appears at $v_R \sim O(10^{15})~{\rm GeV}$
where large deviation is given by $m_{d}~(\sim 1.3\sigma~{\rm smaller})$ and $m_{s}~(\sim 2\sigma~{\rm larger})$.
For type II, the $\chi^2$ curve becomes flat above $v_R \sim O(10^{14})~{\rm GeV}$
and no prominent local minimum exists.
As we expected in Section 3, $v_R \sim O(10^{16})~{\rm GeV}$ is disfavored for type I 
while type II gives a $\chi^2$ value of $\sim 2$\,--\,$2.5$
where the deviation is dominated by the down quark mass ($\sim 1.4\sigma$ smaller).
Compared with the case of $\theta_{23} = 0.710$, although the difference is not obvious, 
$\theta_{23} = \pi/4$ gives slightly larger $\chi^2$ values and shifts the minima to lower energy scale.
Tables \ref{table:fit_typeI} and \ref{table:fit_typeII} show the summary of the fit parameters and $\chi^2$  
at each global best fit ($v_R \sim O(10^{13})~{\rm GeV}$), the local minimum of type I ($v_R \sim O(10^{15})~{\rm GeV}$)
and $R = 0.001$ of type II ($v_R \sim O(10^{16})~{\rm GeV}$).

The numerical quantities of the right-handed neutrino mass are also obtained from the fit.
For type II, we find that the sign combination of $\sqrt{K}$ ($s_{1}, s_{2}, s_{3}$) becomes a permutation of $(-1, 1, 1)$
at a low $R$ region ($R < 0.1$).
Thus, as investigated in section 3, the lightest right-handed neutrino mass is small 
compared with the other two eigenvalues.
For example, at $R = 0.001$, the lightest mass is $6.9~(5.5) \times 10^{8}~\mathrm{GeV} $
while the other masses are 
$3.3~(3.5) \times 10^{14}~\mathrm{GeV} $ and $1.3~(1.2) \times 10^{16}~\mathrm{GeV} $ for 
$\theta_{23} = 0.71~(\pi/4)$.
In addition, the lightest right-handed neutrino mass is insensitive to $R$
while the other two eigenvalues of the neutrino mass behave as $\sim 1/R$ in that small $R$ region, 
which is also consistent with the analytical consideration in section 3.
Around the global best fit ($v_R \sim 10^{13}~{\rm GeV}$), in contrast, 
the lightest neutrino mass grows and becomes nearly the same scale as the second largest neutrino mass. 
For example, at the global best fit $R$, 
the lightest neutrino mass is $2.4~(2.2) \times 10^{10}~\mathrm{GeV} $
while other two eigenvalues are
$4.6~(3.2) \times 10^{10}~\mathrm{GeV} $ and $11.1~(6.9) \times 10^{11}~\mathrm{GeV} $ for 
$\theta_{23} = 0.71~(\pi/4)$.

We assume the normal hierarchy of the neutrino masses, $m_1< m_2<m_3$, in the above discussion. 
We also search for $\chi^2$ minimum for the inverted hierarchy case, which can be done since the
neutrino mass matrix is given as input in our formula. We find that 
the fit does not lead to a competitive result within the energy scale from $10^{13}~{\rm GeV}$ to $10^{16}~{\rm GeV}$,
which gives $\chi^2 > 200$.
\begin{table}[p]
\begin{center}
\begin{tabular}{|c|c|c|c|c|}
\hline
\multicolumn{5}{|c|}{ Type I } \\
\hline
\hline
 & \multicolumn{2}{|c|}{Best fit} & \multicolumn{2}{|c|}{Local minimum} \\ \hline
$\theta_{23}$                       &$  0.710    $&  $\pi/4$    &$  0.710    $&$  \pi/4    $\\ \hline
$ R       $                         &$  2.9853   $&$  4.7315   $&$  0.23713  $&$  0.29006  $\\ \hline
$\alpha_{e}$                        &$  1.2200   $&$ -0.15753  $&$  0.97741  $&$  1.5312   $\\
$\alpha_{\mu}$                      &$ -0.77358  $&$  1.9305   $&$  1.7110   $&$  0.95362  $\\
$\alpha_{\tau}$                     &$  2.2100   $&$ -2.1821   $&$ -1.7127   $&$ -2.3612   $\\
$\alpha_{2}$                        &$ -2.8818   $&$  2.8578   $&$ -2.7599   $&$ -2.8910   $\\
$\alpha_{3}$                        &$ -0.45528  $&$  1.8924   $&$ -1.1936   $&$ -2.0470   $\\
$\delta_{{\rm PMNS}}$               &$ -0.80638  $&$  1.9094   $&$ -1.5819   $&$ -2.3119   $\\
$\log_{10} (m_{1}/{\rm GeV})$       &$-11.501    $&$-11.273    $&$-11.381    $&$-11.270    $\\ \hline
$m_{c}~({\rm GeV})$                 &$  0.1955   $&$  0.1999   $&$  0.2153   $&$  0.2098   $\\
$m_{d}~({\rm GeV})$                 &$  0.0004810$&$  0.0004270$&$  0.0004380$&$  0.0004416$\\
$m_{s}~({\rm GeV})$                 &$  0.01814  $&$  0.01798  $&$  0.02705  $&$  0.02791  $\\
$m_{b}~({\rm GeV})$                 &$  0.9916   $&$  0.9885   $&$  0.9789   $&$  1.004    $\\
$     V_{uc}$                       &$  0.2242   $&$  0.2241   $&$  0.2243   $&$  0.2245   $\\
$     V_{sb}$                       &$  0.003703 $&$  0.003674 $&$  0.004025 $&$  0.004011 $\\
$     V_{ub}$                       &$  0.03692  $&$  0.03692  $&$  0.03693  $&$  0.03612  $\\
$\delta_{{\rm KM  }}$               &$  1.218    $&$  1.189    $&$  1.206    $&$  1.190    $\\ \hline
\multicolumn{5}{|c|}{Pull} \\ \hline
$m_{c}$               &$  0.099 $&$  0.275 $&$  0.893  $&$  0.672 $\\
$m_{d}$               &$ -1.186 $&$ -1.328 $&$ -1.299  $&$ -1.290 $\\
$m_{s}$               &$  0.094 $&$  0.061 $&$  1.876  $&$  2.048 $\\
$m_{b}$               &$  0.061 $&$ -0.042 $&$ -0.362  $&$  0.464 $\\
$     V_{uc}$         &$  0.086 $&$  0.033 $&$  0.156  $&$  0.229 $\\
$     V_{sb}$         &$  0.006 $&$ -0.057 $&$  0.722  $&$  0.691 $\\
$     V_{ub}$         &$ -0.065 $&$ -0.064 $&$ -0.051  $&$ -0.675 $\\
$\delta_{{\rm KM}}$   &$  0.190 $&$  0.045 $&$  0.128  $&$  0.048 $\\
\hline
$ r               $      &$ 0.0138 $&$ 0.0144  $&$ 0.0279  $&$ 0.0285 $\\
$ r_{2}           $      &$ 0.323 + 0.00618 i  $&$ 0.311 - 0.00331 i  $&$ 2.78 + 0.00190 i $&$ 2.77 - 0.379 i $\\
$ c_Rv_R~({\rm GeV})  $ & 
$ 2.85 \times 10^{13} $ & 
$ 1.76 \times 10^{13} $ & 
$ 4.69 \times 10^{14} $ &
$ 3.81 \times 10^{14} $ \\
$\chi^2$              &$  1.48  $&$  1.85  $&$  6.70  $&$  7.51  $\\
\hline
\end{tabular}
\caption{\sl \small 
The fit result for type I. Pull is defined by  
$(\chi_i - \hat\chi_i)/\hat \sigma_i$ for each observable.
}
\label{table:fit_typeI}
\end{center}
\end{table}

\begin{table}[p]
\begin{center}
\begin{tabular}{|c|c|c|c|c|}
\hline
\multicolumn{5}{|c|}{ Type II } \\
\hline
\hline
 & \multicolumn{2}{|c|}{Best fit}            & \multicolumn{2}{|c|}{ $R = 0.001$ }     \\ \hline
$\theta_{23}$                  &$  0.710    $&$  \pi/4    $&$  0.710    $&$  \pi/4    $\\ \hline
$ R $                          &$  5.1582   $&$  7.2861   $&$  0.001    $&$  0.001    $\\ \hline
$\alpha_{e}$                   &$ -0.81968  $&$ -1.8785   $&$ -0.66648  $&$ -0.25301  $\\
$\alpha_{\mu}$                 &$ -1.0015   $&$  1.5169   $&$ -2.8148   $&$  2.8177   $\\
$\alpha_{\tau}$                &$  1.4632   $&$  2.8853   $&$ -0.53961  $&$ -0.84287  $\\
$\alpha_{2}$                   &$ -2.8481   $&$  2.8306   $&$ -2.8709   $&$ -3.1146   $\\
$\alpha_{3}$                   &$ -0.30601  $&$  1.1731   $&$ -1.9809   $&$ -2.8604   $\\
$\delta_{{\rm PMNS}}$          &$ -0.28115  $&$  1.4476   $&$ -2.3550   $&$ -3.1131   $\\
$\log_{10} (m_{1}/{\rm GeV})$  &$-11.592    $&$-11.467    $&$-11.207    $&$-11.173    $\\
$|\delta|$                     &$ 75.056    $&$100.11     $&$ 15.545    $&$ 16.156    $\\
${\rm Arg(\delta)}$            &$  0.11782  $&$  0.11535  $&$  0.43912  $&$  0.51567  $\\ \hline
$m_{c}~({\rm GeV})$            &$  0.1931   $&$  0.1929   $&$  0.1978   $&$  0.1989   $\\
$m_{d}~({\rm GeV})$            &$  0.0009278$&$  0.0009309$&$  0.0004138$&$  0.0003936$\\
$m_{s}~({\rm GeV})$            &$  0.01785  $&$  0.01767  $&$  0.01980  $&$  0.02028  $\\
$m_{b}~({\rm GeV})$            &$  0.9897   $&$  0.9898   $&$  0.9903   $&$  0.9901   $\\
$     V_{uc}$                  &$  0.2240   $&$  0.2240   $&$  0.2240   $&$  0.2241   $\\
$     V_{sb}$                  &$  0.003698 $&$  0.003698 $&$  0.003765 $&$  0.003724 $\\
$     V_{ub}$                  &$  0.03700  $&$  0.03699  $&$  0.03694  $&$  0.03695  $\\
$\delta_{{\rm KM  }}$          &$  1.180    $&$  1.180    $&$  1.195    $&$  1.160    $\\ \hline
\multicolumn{5}{|c|}{Pull} \\ \hline
$m_{c}$               &$   0.004 $&$   -0.005 $&$    0.191 $&$    0.236 $\\
$m_{d}$               &$  -0.010 $&$   -0.002 $&$   -1.363 $&$   -1.416 $\\
$m_{s}$               &$   0.036 $&$    0.000 $&$    0.426 $&$    0.522 $\\
$m_{b}$               &$  -0.002 $&$    0.000 $&$    0.017 $&$    0.010 $\\
$     V_{uc}$         &$   0.002 $&$   -0.012 $&$   -0.000 $&$    0.027 $\\
$     V_{sb}$         &$  -0.005 $&$   -0.005 $&$    0.144 $&$    0.054 $\\
$     V_{ub}$         &$  -0.004 $&$   -0.009 $&$   -0.044 $&$   -0.039 $\\
$\delta_{{\rm KM}}$   &$  -0.001 $&$   -0.001 $&$    0.075 $&$   -0.099 $\\
\hline
$ r               $      &$ 0.0140           $&$ 0.0138            $&$ 0.0230         $&$ 0.0233 $\\
$ r_{2}           $      &$ 0.506 + 0.0252 i $&$ 0.502 + 0.00628 i $&$ 2.15 + 0.227 i $&$ 2.22 - 0.160 i  $\\
$ c_Rv_R~({\rm GeV})   $ & 
$ 1.19  \times 10^{13} $ & 
$ 0.861 \times 10^{13} $ & 
$ 8.86  \times 10^{16} $ &
$ 9.22  \times 10^{16} $ \\
$\chi^2$              &$ 0.001  $&$ 0.0003 $&$ 2.10  $&$ 2.35   $\\
\hline
\end{tabular}
\caption{\sl \small 
The fit result for type II. }
\label{table:fit_typeII}
\end{center}
\end{table}

\newpage

\subsection{Prediction of PMNS phase in the neutrino oscillation}

\begin{figure}[t]
\begin{center}
\includegraphics[width=0.5\textwidth]{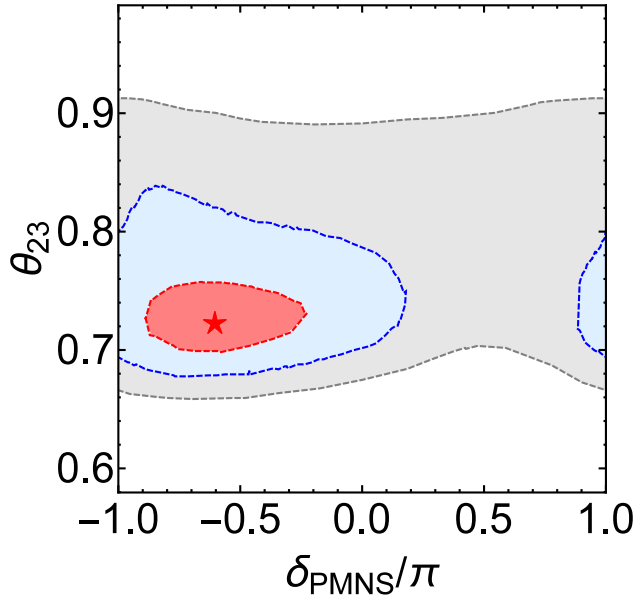}
\caption{\sl \small 
The preferred region of $\delta_{{\rm PMNS}}$ and $\theta_{23}$ 
given by~\cite{Capozzi:2013csa}.
The best fit is shown by the red star.
The red, blue, gray area represents $1\sigma,\, 2\sigma$ and $3\sigma$
region respectively.
}
\end{center}
\label{fig:ddd}
\end{figure}

The PMNS phase $\delta_{{\rm PMNS}}$ is one of the interesting predictions of our fitting results.
Though the accuracy is not still sufficient, 
the on-going experiments~\cite{Wilking:2013vza} already give
a tendency of the parameter region.
The long-baseline accelerator experiments and its global fit~\cite{Elevant:2015ska}
suggest that $\theta_{23}$ larger than $\pi/4$ is slightly favored
while the reactor experiments support small $\theta_{23} (< \pi/4)$.
Fig.~2 shows the global best fit by~\cite{Capozzi:2013csa}
which involves both the accelerator and the reactor experiments\footnote{
The global best fit is recently updated by~\cite{Capozzi:2016rtj}.
The result does not significantly changed.
}.
Although the wide range of the parameter region is still allowed,
negative $\delta_{{\rm PMNS}}$ with small $\theta_{23}$ ($<\pi/4$) is slightly favored. 
It is expected that the experiments  will present 
more accurate $\delta_{{\rm PMNS}}$ as well as $\theta_{23}$ within a few years. 
In addition, accessibility of other future experiments is estimated 
and found to reach 
ten and several percent accuracy for $\theta_{23}$ and
10 to 40 percent accuracy for $\delta_{{\rm PMNS}}$ \cite{Coloma:2014kca}. 
Therefore, it is interesting to compare the current experimental favored region with our fit prediction.

For that purpose, we search for the global $\chi^2$ minimum for each fixed $\theta_{23}$ and $\delta_{{\rm PMNS}}$\footnote{
One might think that we can construct the same map by 
simply accumulating the MCMC samples by allowing $\theta_{23}$ and $\delta_{{\rm PMNS}}$ free.
However, as mentioned in section 4.2, the $\chi^2$ function has a large number of deep local minimum
and it is difficult to explore the detailed $\chi^2$ map under the full free parameter space.
Hence, we decide to fix these two parameter for each MCMC simulation.
}
in the range of $ 0.675 < \theta_{23} < 0.835$ and 
$ |\delta_{{\rm PMNS}}| < \pi$ allowing $R$ free within a range of $0.0001 < R < 100$
 and also explore the same parameter region at $R = 0.001$ ($ c_{R}v_{R} \sim 10^{16}~{\rm GeV}$) for Type II. 
Here we set the parameter range of $\theta_{23}$ as $2\sigma$ region of Ref.~\cite{Capozzi:2013csa}.

\begin{figure}[t]
\begin{center}
\includegraphics[width=0.30\textwidth]{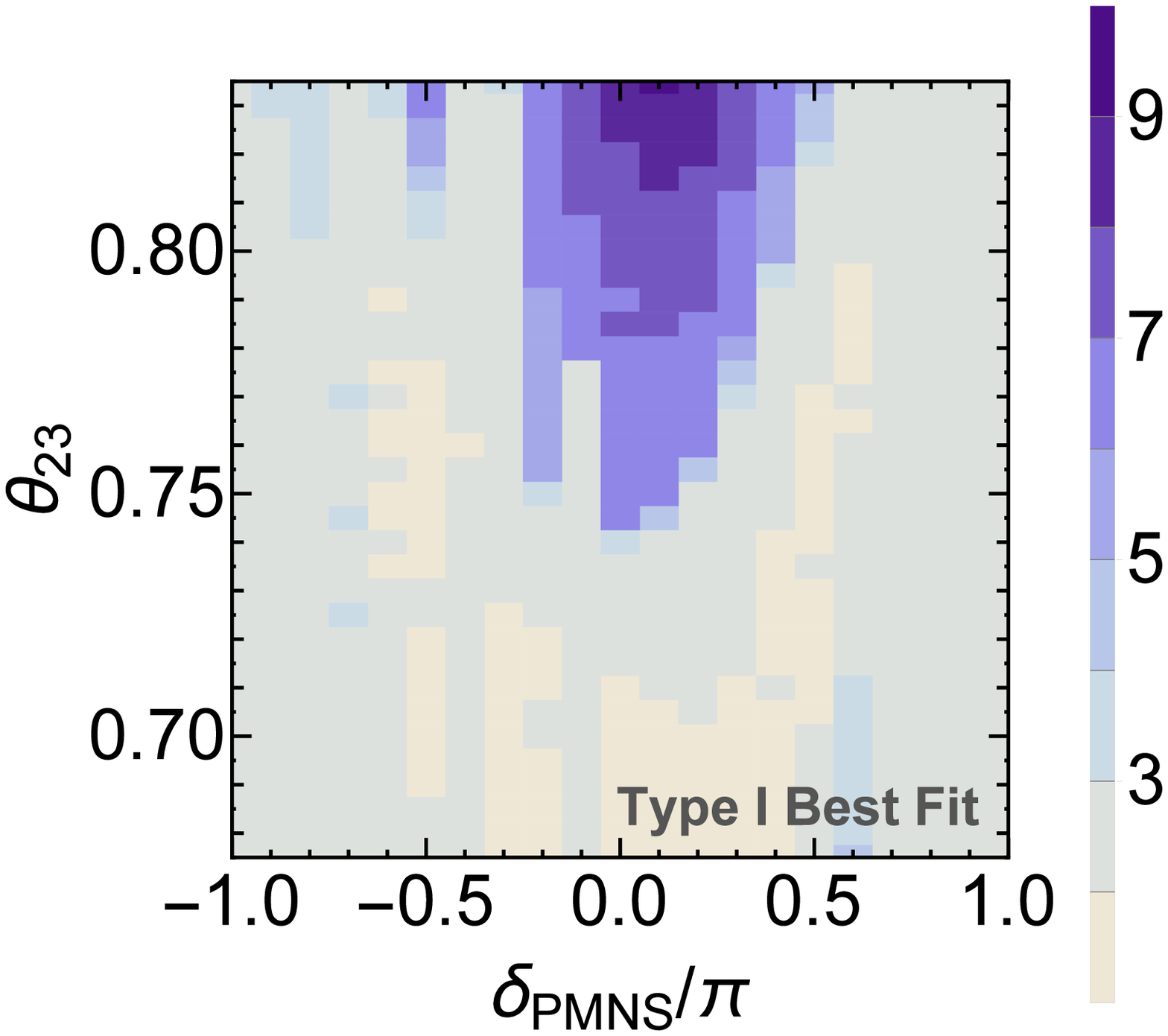}
\includegraphics[width=0.31\textwidth]{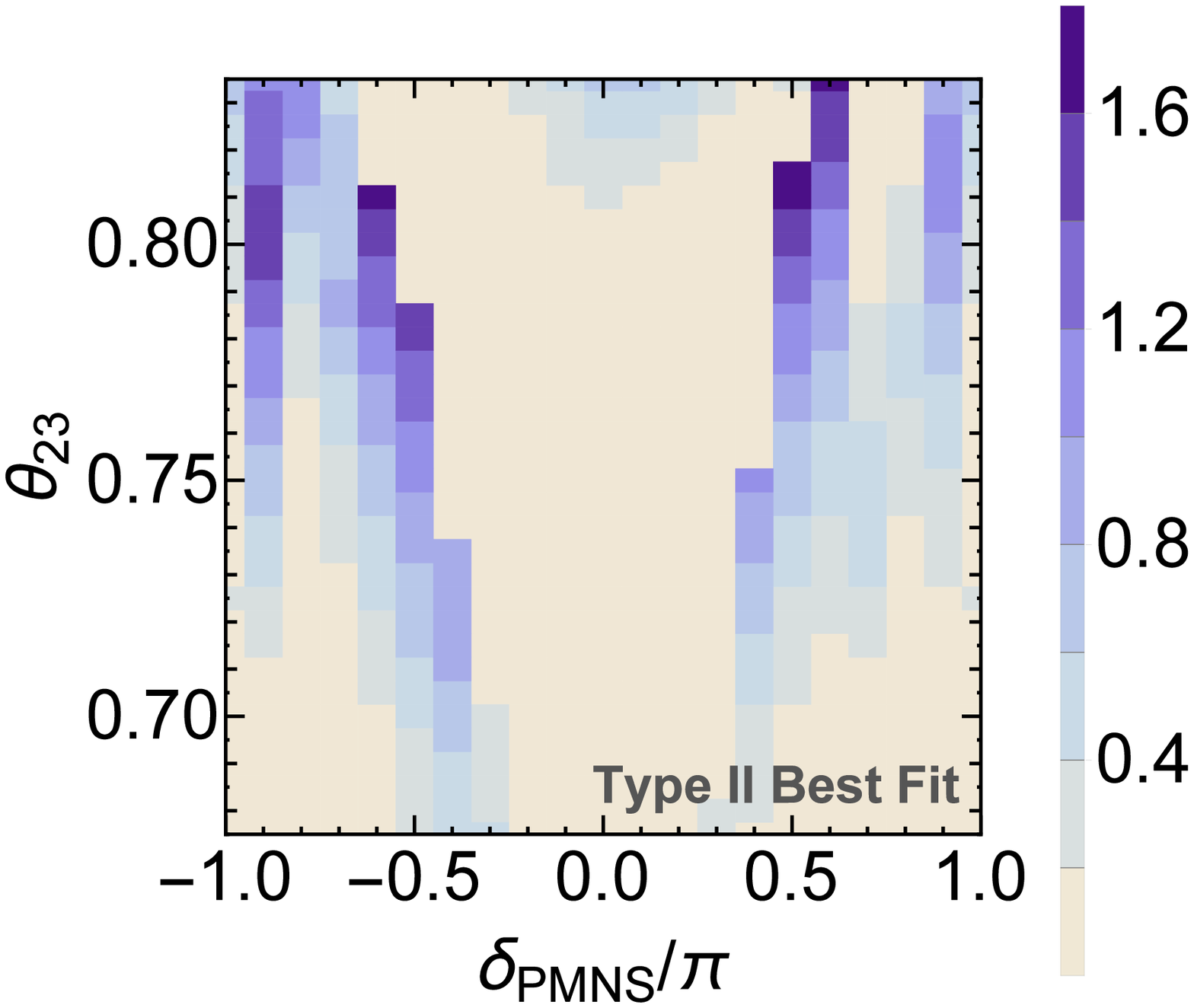}
\includegraphics[width=0.32\textwidth]{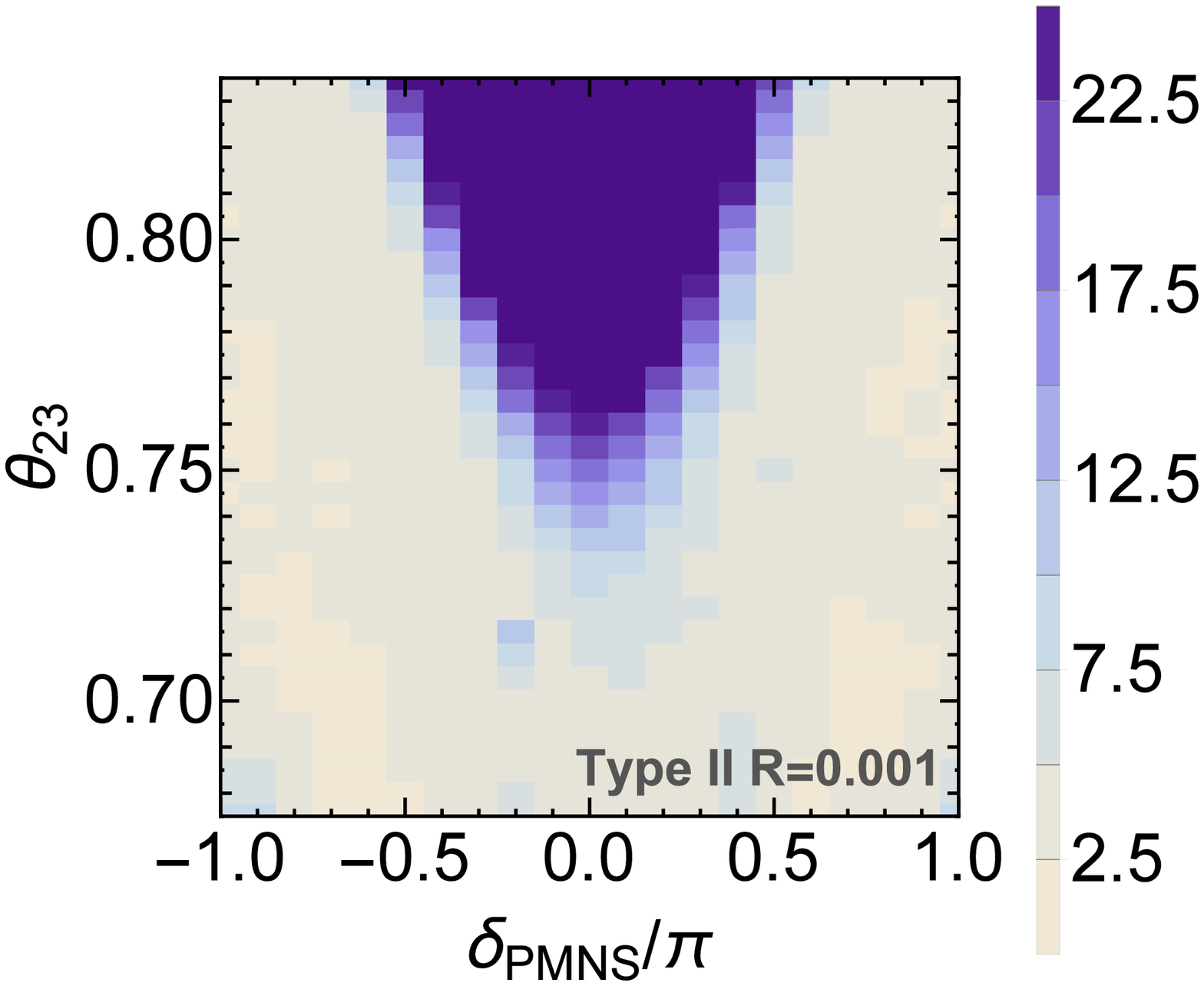}
\caption{\sl \small 
{\bf Left panel:}
The $\chi^2$ map of type I global best fit.
The colored tiles show the magnitude of $\chi^2$.
{\bf Middle panel:}
The $\chi^2$ map of type II global best fit.
{\bf Right panel:}
The $\chi^2$ map of type II best fit at $R = 0.001$.
We set the darkest blue panel to represent $\chi^2 > 25$.
}
\end{center}
\label{fig:del0}
\end{figure}

Fig.~3 shows the $\chi^2$ map with respect to $\delta_{{\rm PMNS}}$ and $\theta_{23}$ for each case. 
The darker blue region represents higher $\chi^2$
where the step size is given by the color bars beside the map.
For type I global best fit (left panel), the figure shows that 
$|\delta_{{\rm PMNS}}/\pi| < 0.3$ is mildly disfavored when $\theta_{23}$ is large $( > \pi/4 )$.
At most of the points, the largest contribution to $\chi^2$ comes from (smaller) $m_{d}$ with $1$\,--\,$2.5\sigma$
deviation.
The second largest contribution is from (larger) $m_{s}$ and (smaller) $\delta_{\mathrm{KM}}$,
which becomes conspicuous at the disfavored region and reach $1$\,--\,$1.5\sigma$.
Interestingly, the $\chi^2$ asymmetry on the $\delta_{{\rm PMNS}}$ sign is mainly derived 
from the $\delta_{\mathrm{KM}}$ deviation,
the center of which is located at around  $\delta_{{\rm PMNS}}/\pi \sim 0.1$.
If $\delta_{\mathrm{KM}}$ is more precisely measured in the future, this asymmetry will be stressed, 
which leads a stronger prediction on the sign of the PMNS phase.

For type II, the $\chi^2$ map of 
the global best fit (middle panel) shows that 
$\chi^2$ is small enough under any value of $\theta_{23}$ and $\delta_{{\rm PMNS}}$.
In contrast to the type I case,
larger $\chi^2$ appears at $|\delta_{{\rm PMNS}}/\pi| > 0.5$ in a large $\theta_{23}\,(>0.75)$ region,
implying that the fit becomes an interesting prediction
if $\theta_{23}$ is large as suggested by the current accelerator experiments.
The largest contribution to $\chi^2$ is from $m_{d}$ ($0$\,--\,$1.2\sigma$ smaller)
and the second largest contribution is from $m_{s}$ ($0$\,--\,$0.7\sigma$ smaller)
and the other contributions are negligible.
We also check the $\chi^2$ behavior at large $\theta_{23}$ ($0.835 < \theta_{23} < 0.88$) 
under fixed $\delta_{{\rm PMNS}}/\pi = -0.5$
and find that the $\chi^2$ value remains less than 3.0 for all the $\theta_{23}$ region.
We expect that more accurate measurement against the observables used in our fit 
will resolve the degeneracy of the $\chi^2$ map in the near future.

For type II of $ R = 0.001 $ (right panel), 
though the shape of the $\chi^2$ map is similar to the type I best fit,
it seems more symmetric with $\delta_{{\rm PMNS}}$-axis and 
its disfavored region is more obvious.
The $\chi^2$ value within $|\delta_{{\rm PMNS}}/\pi| < 0.5$ steeply increases as $\theta_{23}$ becomes large 
and reaches $\sim 80$ at $\delta_{{\rm PMNS}}/\pi \sim 0$.
(In the figure, we set the darkest blue panel to represent $\chi^2 > 25$.)
Particularly, 
$\delta_{{\rm PMNS}}/\pi \sim -0.5$ with a large $\theta_{23}\,(>0.81)$ 
is strongly disfavored in this phenomenologically preferred case.
Stronger upper bound on $\theta_{23}$ can be found
when $|\delta_{{\rm PMNS}}|$ is small
and reaches $\theta_{23} < 0.73$ at $ |\delta_{{\rm PMNS}}/\pi| \sim 0$.
At a point with a small $\chi^2\,(<5)$ in our map,
the largest contribution to $\chi^2$ comes from $m_{d}$ ($1$\,--\,$1.5\sigma$ smaller) 
and $m_{s}$ ($0$\,--\,$1.5\sigma$ larger).
While the $m_{d}$ contribution does not strongly depend on $\theta_{23},\,\delta_{{\rm PMNS}}$,
the $m_{s}$ deviation becomes large at the disfavored region 
($|\delta_{{\rm PMNS}}/\pi| < 0.5$ with a large $\theta_{23}$)
and reaches $\sim 2.5\sigma$.
Still, these contributions are subdominant at the disfavored region 
in which (larger) $m_{c}$ and (smaller) $\delta_{\mathrm{KM}}$ give
$4$\,--\,$6 \sigma$ and $3$\,--\,$4.5 \sigma$ deviations, respectively.
In addition, (larger) $m_{b}$ and (larger) $V_{ub}$ also contribute to $\chi^2$ with
$3$\,--\,$4 \sigma$ deviation
around $|\delta_{{\rm PMNS}}/\pi| \sim 0$ at a large $\theta_{23}\,(> \pi/4)$.
Similarly to the type I case, 
the $\delta_{\mathrm{KM}}$ deviation is slightly $\delta_{{\rm PMNS}}$-axis asymmetric,
which is centered at $ |\delta_{{\rm PMNS}}/\pi| \sim 0.1$.
However, this asymmetry is canceled by the contribution from 
$m_{c},\,m_{b}$ which are located at around $ |\delta_{{\rm PMNS}}/\pi| \sim -0.1$. 
Hence, we cannot conclude which sign of $\delta_{{\rm PMNS}}$ is preferred only from our fit.

\begin{figure}[t]
\begin{center}
\includegraphics[width=0.5\textwidth]{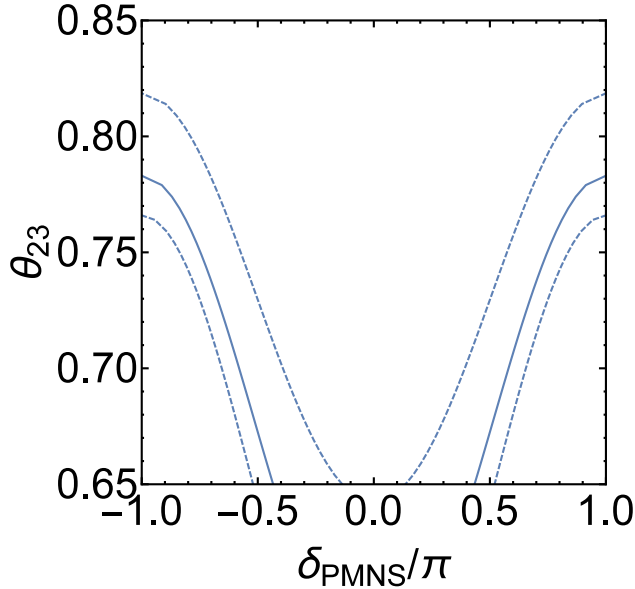}
\caption{\sl \small 
The relation between $\delta_{\rm PMNS}$ and $\theta_{23}$
under the two-zero texture assumption: $({\cal M}_\nu)_{11} = ({\cal M}_\nu)_{12}=0$.
The solid line is drawn using the center value of $\theta_{13}$ : $\sin\theta_{13} = 0.156$,
and the dashed lines are drawn to present the $2\sigma$ range of $\theta_{13}$.
For the mass squared difference and $\theta_{12}$, we use the center values of the global analysis \cite{Capozzi:2013csa}.
}
\end{center}
\label{fig:PMNS}
\end{figure}

Even though it is not easy to explain how the prediction of the phase is obtained explicitly,
one can understand the results qualitatively, especially in the case of $R=0.001$ for type II.
Relating to the result that the fit of the down quark mass is smaller than the observation,
(1,1) and (1,2) elements of ${\cal M}_\nu$ are smaller than the other elements in the fit result.
In fact, in the special case with two-zero texture of the neutrino mass matrix, $({\cal M}_\nu)_{11} = ({\cal M}_\nu)_{12}=0$,
one obtains a relation among the mixing angles, mass squared ratio, and PMNS phase \cite{Fritzsch:2011qv}. 
Under the two-zero texture assumption, we obtain \cite{Haba:2011pe,Dutta:2013bvf}
\begin{equation}
\cos\delta_{\rm PMNS} = 
\frac{ \frac{\Delta m_{\rm sol}^2}{\Delta m_{\rm atm}^2} \cos2\theta_{13}\sin^22\theta_{12}   - 
4\sin^2\theta_{13} \left(\frac{\Delta m_{\rm sol}^2}{\Delta m_{\rm atm}^2} \cos^4\theta_{12}+\cos2\theta_{12}\right)
\tan^2\theta_{23}}{4\sin^3\theta_{13} \left(1+  \frac{\Delta m_{\rm sol}^2}{\Delta m_{\rm atm}^2} \cos^2\theta_{12}\right)\sin2\theta_{12}\tan\theta_{23}}.
\end{equation}
In Fig.4, we plot the relation between $\delta_{\rm PMNS}$ and $\theta_{23}$ in the assumption.
Surely, those two elements are not exactly zero in the fits, and the relation provide a guide to understand the fit results
for the prediction of the PMNS phase depending on the mixing angle.
In fact, the points far from the two-zero texture, such as $(\delta_{\rm PMNS}, \theta_{23}) \sim (0, 0.8)$,
are disfavored in our fits in the case of $R=0.001$ for type II.
There is also such a tendency in the fits in type I, though the detail of the fits is more complicate.
In the type II global best fit, on the other hand, the (1,1) and (1,2) elements are not small
for the observed values to be tuned.
Therefore, the $\chi^2$ values become slightly larger near the points from the two-zero texture.
We remark that the smallness of (1,1) and (1,2) elements relates to the suppression of the proton decay amplitudes \cite{Dutta:2004zh}.
It is important to measure the PMNS phase and the 23-mixing angle accurately by the long-baseline neutrino oscillation experiments
in order to survey the structure of the Yukawa matrices in the $SO(10)$ model.

\section{Implication of the scale $v_R$}

We have demonstrated the fitting of the fermion masses and mixings
to make the properties of solution depending on $v_R$ conspicuously.
As we have described, the key to distinguish the scale $v_R$ in the model 
is the combination of the precise measurements of 23-mixing angles and PMNS phases.
Surely, the fits are shown by making the $\chi^2$ analyses, and the prediction of the PMNS phase 
is modest by referring the $\chi^2$ values in the parameter space, though the fits will be more crucial
when the parameters are measured more accurately in the future.
Not only the numerical values of the observables but also the 
flavor structure in the Yukawa matrices are different for each 
solution as we have displayed in the previous section.
The detail of the flavor structure can be applied to the low energy phenomena, which can depends on the scale $v_R$.
In this section, we study the application,
which can depends on the flavor structure in the several cases of the model.

\subsection{Lepton flavor violation}

The fitting of the fermion masses and mixings in the minimal $SO(10)$ model 
fixes the matrices of the Yukawa couplings.
The most interesting application of this feature is
to calculate the flavor violation in the SUSY model.
In fact, the Dirac and Majorana neutrino mass matrices are determined
(as a function of $v_R$ scale),
and the flavor violation is predictable.
The importance of the scale $v_R$ can be also illustrated 
in the scheme where the SUSY breaking mass matrices are diagonal as boundary conditions
and the flavor violation is generated only from the Yukawa interaction via RGE running.
As an application of the solution of the mass and 
mixing fitting, we perform
 the calculation of lepton flavor violation of $\mu\to e\gamma$ process 
by assuming the minimal
flavor violation, which can be a major sensitive probe to find a footprint.

There are two kinds of amplitudes of $\mu\to e\gamma$ decay,
$A_L : \mu_R \to e_L \gamma$ and $A_R : \mu_L \to e_R \gamma$,
and the branching fraction of $\mu \to e\gamma$ is 
\begin{equation}
{\rm Br} (\mu\to e\gamma) \propto |A_L|^2 + |A_R|^2.
\end{equation}
Roughly speaking,
$A_L$ arises from the chargino loop contribution and the off-diagonal elements in the left-handed slepton mass matrix
contribute to it.
The right-handed amplitude arises from neutralino loop diagrams :
(1) Bino-Higgsino diagram, (2) Bino-Bino diagram.
The off-diagonal elements in the right-handed charged lepton mass matrix contribute to the diagram (1),
and those of both left- and right-handed slepton mass matrices can contributes to the diagram (2).

The sources of the flavor violation are the following:
\begin{enumerate}
\item
Dirac Yukawa coupling, $\ell \nu^c H_u$.

The right-handed neutrino $\nu^c$ and the Higgs doublet $H_u$ propagate in the loop,
and the off-diagonal elements of the left-handed slepton mass matrix are generated \cite{Borzumati:1986qx}.
In the minimal $SO(10)$ model, the Dirac neutrino Yukawa coupling and the heaviest right-handed neutrino mass
are determined (irrespective of $v_R$), and thus, this contribution is predictable.

\item
Majorana coupling, $\ell \ell \Delta_L$.

The $SU(2)_L$ triplet $\Delta_L$ and the left-handed lepton doublet $\ell$ propagate in the loop,
and the off-diagonal elements of the left-handed slepton mass matrix are generated.
In the triplet seesaw, when the $SU(2)_L$ triplet contribution is in the sub-eV range,
the mass of the $\Delta_L$ is about $10^{13}$ GeV, and this contribution will be important to the lepton flavor violation.
Since the coupling matrix is unified to the $\overline{\bf 126}$ Higgs coupling,
the Majorana coupling matrix is determined by the fermion mass and mixing fitting, up to the overall mixing parameter.
The coupling matrix $f$ is obtained as the original matrix $\bf f$ multiplied by the Higgs mixing as given in section 2,
and thus, this contribution is not fully determined by the fit of the fermion masses and mixings.
Since the matrix is obtained up to the over all factor,
the ratio of the contributions to $\mu\to e\gamma$, $\tau\to\mu\gamma$ and $\tau \to e\gamma$ is predictable
if the contributions from the other sources are small.
Because the Higgs mixing cannot be larger than 1, the matrix elements of the Majorana coupling
have lower bounds in the simplest model.
However, if the ${\bf 54}$ Higgs is adopted, the $SU(2)_L$ triplet from the $\overline{\bf 126}$ 
is mixed with the one in ${\bf 54}$,
and then, the $SU(2)_L$ triplet mixing is also multiplied, and the lower bound from this source is model-dependent.

\item
$e^c e^c \Delta_R^{--}$, $e^c \nu^c \Delta_R^-$ couplings.

If the $SU(2)_R$ symmetry or $U(1)_R \times U(1)_{B-L}$ symmetry is broken
much below the GUT scale,
the fields of $\Delta_R^{--}$ and/or $\Delta_R^-$ can be light
and they propagate in the loop
and the off-diagonal elements of the right-handed charged-lepton mass matrix is generated.
This contribution is related to the Higgs spectrum and model-dependent,
and thus, it is not very predictable.
However, if $v_R$ is $\sim O(10^{13})$ GeV, this contribution is not small
and need to be considered.
Since it contributes to the FCNC from the right-handed sleptons,
flavor violating decay is generated from the neutralino loop diagram.

\item
$e^c u^c H_C$ coupling.

In the language of $SU(5)$ GUT, 
${\bf 10}\cdot {\bf 10} \cdot {\bf 5}_H$ includes this coupling.
The right-handed up-type quarks and the colored Higgs fields $H_C$ propagate in the loop,
and the off-diagonal elements of the right-handed charged-lepton mass matrix are generated.
In the other sources, the mass of the fields which propagate in the loop is supposed to be of the order of $10^{13}$ GeV,
while the colored Higgs mass is $O(10^{16})$ GeV,
and thus, this contribution is not very important, compared to the others in the current context of $SO(10)$ model.

\end{enumerate}

In general, those Yukawa coupling $Y_{ij}$ to the charged-leptons can induce off-diagonal elements in the slepton mass matrices
by RGE in the form of
\begin{equation}
(M_{\tilde \ell,\tilde e}^2){}_{i\neq j} = - \frac{C}{8\pi^2} \sum_k Y_{ik} Y^*_{jk} (3 m_0^2 + A_0^2) \ln \frac{M_*}{M_X}.
\end{equation}
Here $m_0$
is a universal scalar mass, $A_0$ is a universal scalar trilinear coupling,
$M_*$ is a cutoff scale, $M_X$ is the mass of a heavy field which propagate in the loop,
and $C$ is a group weight factor.
In particular, the off-diagonal elements induced by the 
Dirac neutrino Yukawa coupling $Y_\nu$ 
can be roughly written as
\begin{equation}
(M_{\tilde \ell}^2){}_{i\neq j} = - \frac{1}{8\pi^2} \sum_k(Y_\nu)_{ik} (Y_\nu)^*_{jk} (3 m_0^2 + A_0^2) \ln \frac{M_*}{M_{R_k}},
\end{equation}
where $Y_\nu$ is given in the basis
where the right-handed neutrino mass matrix is diagonal,
and 
$M_{R_k}$ is an eigenmass of $k$-th generation right-handed neutrino.
It is important to note that
the gluino and squark masses should be heavy due to the LHC results,
and thus, the amount of the induced FCNC becomes less in the universal SUSY breaking models.
The discovery of the 125 GeV Higgs boson also pushes up the squark masses.
If the mass of the squarks and gluino are O(10) TeV, it is hard to extract the off-diagonal elements from the flavor data.
However, if the squark and gluino masses are about 2 TeV, the scalar trilinear coupling $A_0$ has to be large ($\sim$ 5 TeV)
to obtain the Higgs mass to be 125 GeV,
and then, the off-diagonal elements are generated (even if $m_0$ is small)
and the FCNCs are induced slightly.
Therefore, the SUSY contribution can be consistent with the experimental results of many of the FCNC processes,
but a slight excess can be observed in a process whose amplitude can have an enhancement factor.
We remark that the circumstances are changed from the literatures a few years ago.

Other main circumstance changed by the LHC experiments is the SUSY mass spectrum:
The gluino and squark masses are bounded from below.
The observation of $B_s \to \mu^+\mu^-$ to be consistent with the SM prediction also bounds the parameters \cite{CMS:2014xfa}.
Though the bound of the slepton masses, which is important for the lepton flavor violating decays,
is not directly related to the gluino and squark mass bound,
the slepton masses depend on the gluino and squark mass bounds if the universal SUSY breaking is assumed.
The bounds depend on the SUSY breaking scenario.
Since the anomaly mediation is an infrared phenomenon, the slepton spectrum is still less bounded
compared to the squark masses, even in the GUT models\footnote{
In fact, if we take the known anomaly of the muon $g-2$ seriously,
the universal SUSY breaking is not very favored after the LHC results bound the gluino and squark masses,
and the mixed modulus-anomaly mediation is favored in the unification scenario \cite{Chowdhury:2015rja}.
In that scenario, the branching fraction of $\mu\to e\gamma$ becomes larger
and the flavor non-universality (as we will describe in Section 10) is needed to satisfy the current bound.
}.
We calculate the lepton flavor violation using the Yukawa couplings by fitting 
the fermion masses and mixings in the minimal $SO(10)$ model
assuming the universality to make the number of SUSY breaking parameters less,
and thus, the slepton mass spectrum depends on the gluino mass bound indirectly.
We choose the unified gaugino mass $m_{1/2} = 800$ GeV to satisfy 
the current gluino bound and to be covered the next run at the LHC.
We vary the universal scalar mass $m_0$, which is important to the slepton masses.
The scalar trilinear coupling $A_0$ is chosen to make the Higgs mass to be 125 GeV, depending on $m_0$.
The ratio of the vev of up- and down-type Higgs bosons, $\tan\beta$, is an important parameter
because the decay width of $\mu\to e\gamma$ is roughly proportional to $\tan^2\beta$.
We choose $\tan\beta=10$ in the calculation.

We show the results of the numerical calculation in four cases given by the $\chi^2$ minimum fits,
where the right-handed neutrino masses and the Dirac neutrino Yukawa couplings are given as follows in the basis
where the charged lepton and the right-handed neutrino mass matrices are real/positive diagonal. 
Using unphysical phase freedom in the SM, we make the $(i,1)$ components to be real. 
\begin{enumerate} 
\item
type II, $c_R v_R = 8.86 \times 10^{16}$ GeV ($R=0.001$)
\begin{equation}
M_{R_1} = 6.9 \times 10^8\ {\rm GeV},\quad
M_{R_2} = 3.3 \times 10^{14}\ {\rm GeV},\quad
M_{R_3} = 1.2 \times 10^{16}\ {\rm GeV},
\end{equation}
\begin{equation}
Y_\nu
= \left(
 \begin{array}{ccc}
    0.000111 & 0.000203 + 0.000217\,i & 0.00888 + 0.00372\, i \\ 
    0.000440 & 0.0308 - 0.0248\, i & 0.0426+ 0.0013 \,i \\
    0.00607  & -0.0276 - 0.0069\, i & 0.990 - 0.278\, i
 \end{array}
\right),
\end{equation}

\item
type II, $c_R v_R = 1.19 \times 10^{13}$ GeV ($R=5.1582$) (the best fit)
\begin{equation}
M_{R_1} = 2.4 \times 10^{10}\ {\rm GeV},\quad
M_{R_2} = 4.6 \times 10^{10}\ {\rm GeV},\quad
M_{R_3} = 1.1 \times 10^{12}\ {\rm GeV},
\end{equation}
\begin{equation}
Y_\nu
= \left(
 \begin{array}{ccc}
    0.00301 & -0.00147 - 0.00394 \,i & -0.0031 + 0.0134\, i \\ 
    0.0211 & 0.00023 - 0.0284\, i & -0.0073+ 0.0359 \,i \\
    0.103 & -0.044 - 0.131 \, i & -0.324 + 0.428\, i
 \end{array}
\right),
\end{equation}

\item
type I, $c_R v_R = 4.69 \times 10^{14}$ GeV ($R=0.23713$)
\begin{equation}
M_{R_1} = 6.6 \times 10^{8}\ {\rm GeV},\quad
M_{R_2} = 1.3 \times 10^{12}\ {\rm GeV},\quad
M_{R_3} = 5.5 \times 10^{13}\ {\rm GeV},
\end{equation}
\begin{equation}
Y_\nu
= \left(
 \begin{array}{ccc}
    0.0000957 & 0.000056 - 0.000189 \,i & 0.00852 - 0.00459\, i \\ 
    0.000222 & 0.0279 - 0.0253\, i & 0.0418-0.0143 \,i \\
    0.00400 & -0.0081+ 0.0177\, i & -0.02 - 1.11\, i
 \end{array}
\right),
\end{equation}

\item
type I, $c_R v_R = 2.85 \times 10^{13}$ GeV ($R=2.9853$) (the best fit in type I)
\begin{equation}
M_{R_1} = 6.8 \times 10^{9}\ {\rm GeV},\quad
M_{R_2} = 3.6 \times 10^{11}\ {\rm GeV},\quad
M_{R_3} = 3.0 \times 10^{12}\ {\rm GeV},
\end{equation}
\begin{equation}
Y_\nu
= \left(
 \begin{array}{ccc}
    0.000443 & 0.00236 + 0.00016 \,i & 0.0024 + 0.0109\, i \\ 
    0.00302 & -0.00648 - 0.00218\, i & 0.0105+0.0413 \,i \\
    0.0208 & 0.100 - 0.0004\, i & 0.004 + 0.527\, i
 \end{array}
\right),
\end{equation}

\end{enumerate}

\begin{figure}[t]
\begin{center}
\includegraphics[width=0.55\textwidth]{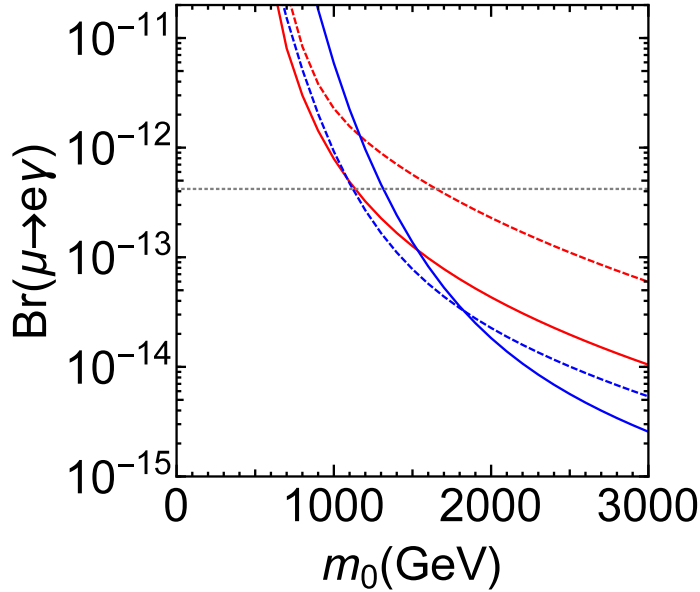}
\caption{\sl \small 
The branching ratio $\mu \rightarrow e \gamma$ for 
type II at $c_Rv_R = 8.86 \times 10^{16}~{\rm GeV}$ (R = 0.001) (red solid line),
type II at the global minimum ($c_Rv_R = 1.19 \times 10^{13}~{\rm GeV}$) (red dashed line),
type I at the local minimum ($c_Rv_R = 4.69 \times 10^{14}~{\rm GeV}$) (blue solid line),
and
type I at the global minimum ($c_Rv_R = 2.85 \times 10^{13}~{\rm GeV}$) (blue dashed line).
We set $\theta_{23} = 0.71$.
 The gray dashed line shows the current experimental bound, Br($\mu \to e\gamma) < 4.2 \times 10^{-13}$ \cite{TheMEG:2016wtm}.
}
\end{center}
\label{fig:bgs}
\end{figure}

As explained, the effects from 
$e^c e^c \Delta_R^{--}$, $e^c \nu^c \Delta_R^-$ couplings
are absent in the case 1,
and 
in type I solutions (3 and 4), the contribution from the Majorana coupling is absent.
The contribution from the left-handed Majorana coupling has an ambiguity from the Higgs mixings,
and here we assume the mixing to make the contributions to the flavor violation maximal.
In the figure, the behavior of the plot in case 3 looks different.
This is because the (3,3) element of the Dirac neutrino Yukawa coupling is a bit larger than others,
and the stop masses are reduced indirectly due to the RGE evolutions,
and a larger value of the $A$-term is needed to reproduce 125 GeV Higgs mass compared to the others.
(The (3,3) element in case 1 is also large, but the effect is not large since the 3rd right-handed neutrino 
is heavy and soon decoupled in the RGE evolution.)
The induced off-diagonal elements in the slepton mass matrices are proportional to $3m_0^2 + A_0^2$,
and thus, the $m_0$ dependence of Br($\mu\to e\gamma$)
looks different between the smaller and the larger $m_0$.
As noted, the value of $A_0$ is chosen to make the Higgs mass 125 GeV,
the detail numerical values depend on the Higgs mass.
We note that due to a large value of $A_0$ ($\sim$ 4-6 TeV), the stau becomes LSP for $m_0 < 600$ GeV.
The branching fraction of the case 2 is the largest (in the region satisfying the current experimental bound).
This is because the (1,1) and (1,2) elements of the Dirac neutrino Yukawa coupling are larger than the other cases,
relating to the result of that the proper size of the down quark mass fit in the case 2.

Relating to the $v_R$ dependence, we here comment on the electron EDM.
We assume $A_0$ and the Higgsino mass $\mu$ to be real because otherwise the EDM becomes much larger than 
the current experimental bound.
Even in the case, if there are $e$-$\tau$ flavor changes (with CP phases) for both left- and right-handed slepton mass matrices,
the Bino-Bino diagram can hit the $A_\tau v_d - \mu m_\tau \tan\beta$ for the electron EDM amplitudes,
and the value of the EDM can be more than the current experimental bound for slepton masses to be $\sim 1$ TeV.
Surely, if there is only left-handed flavor change such as the case of $c_R v_R = 10^{16}$ GeV, the electron EDM is tiny
 as long as $A_0$ and $\mu$ are real.
Therefore, if the scale of $v_R$ is much less than the unification scale, and the $e^c \nu^c \Delta_R^-$ coupling
can generate the right-handed slepton flavor changing, the electron EDM can be large in the $SO(10)$ model,
and the electron EDM can be the probe of the scale $v_R$.

In the numerical calculation in the above cases,
the electron EDM becomes large except for the case 1 for $v_R \sim 10^{16}$ GeV.
In the case 2, the current bound of the electron EDM, $|d_e|< 8.7 \times 10^{-29} \ e \cdot {\rm cm}$ \cite{Baron:2013eja},
 can be satisfied for $m_0> 2.5$ TeV,
 and then, we obtain Br($\mu\to e\gamma) <1.1 \times 10^{-13}$.
In the case 4 (the best fit in type I),
we need $m_0 >2$ TeV, and then Br($\mu\to e\gamma) <2.3 \times 10^{-13}$.
In the case 1, the electron EDM is tiny ($< 10^{-35} \ e\cdot {\rm cm}$)
and $m_0 \simeq 1$ TeV is still allowed, which can provide the boundary value of the current $\mu\to e\gamma$ bound.

\subsection{Proton decay}

As we have explained, the flavor structure is different in each solution,
which influences the proton decay amplitudes depending on the scenarios of seesaw mechanism
and the scale $v_R$.
The calculation of the proton decay amplitudes in various scenarios has been done in literatures \cite{Goh:2003nv},
and we do not repeat it in this paper.
However, in the earliest works, they sometimes neglect right-handed proton decay operators to calculate the decay rate,
and the importance of the flavor structure (especially for the (1,1) and (1,2) elements in the Yukawa matrices) 
for natural suppression of the proton decay amplitudes \cite{Dutta:2004zh}
is missing.
We stress that the flavor structure is important to suppress the proton decay amplitudes in this model,
and claim the importance of the large $v_R$ solution.
We can show that the proton decay amplitudes can be suppressed 
near the line shown in Fig.4 in the solution.
The detail will be discussed in a separate paper \cite{Fukuyama:2016vgi}.

\subsection{Discussion on the modification of Yukawa couplings by threshold corrections}

The quark masses and mixings can receive large corrections from the low energy threshold effects.
This inversely means that the present procedure in the $SO(10)$ model
(inputting the lepton parameters and outputting quark masses and mixings)
provides a prediction to holomorphic Yukawa couplings for quarks at the unification scale,
and the deviations from the observed values of quark masses can be buried by 
the threshold effects \cite{DiazCruz:2000mn,Enkhbat:2009jt,Iskrzynski:2014sba}.
This feature can be applied to suppress the dimension-five operators in the SUSY GUT models.

As given in the previous section,
in the case $R\sim O(1)$ ($v_R \sim O(10^{13})$ GeV),
the down quark mass (and all the other masses and mixings)
can be fully fit.
In the solution, the (1,1) element of the $\overline{\bf 126}$ Higgs coupling
(in the basis where $M_d$ is diagonal)
is as large as the down quark Yukawa coupling.
On the other hand, the solutions for $R \ll 1$,
the (1,1) element can be much less than the down quark Yukawa coupling,
which has a merit to the natural suppression of the proton decay amplitudes
though the fit of the down quark mass deviates from the observation at around 2$\sigma$.

In the SUSY models, in addition to the corrections from the RGEs,
the finite loop corrections arise due to the SUSY breaking.
It is famous that the bottom quark mass can be corrected largely
by the finite corrections of the non-holomorphic term, $q d^c H_u^*$.
Actually, the quark mass matrices, which are connected to the GUT scale via RGEs, 
can be modified by the loop correction via the SUSY breaking around TeV scale,
and in general, all the quark masses and mixings can be different from them via the RGE evolution.
However, 
the corrections from the loop corrections are severely restricted from the flavor physics data,
such as $K$-$\bar K$, $B$-$\bar B$ mixings, $b\to s\gamma$, and $B_{s,d} \to \mu^+\mu^-$,
and there is not so much room to be available.
For example, if the bottom quark mass is modified, the $b\to s\gamma$ and $B_{s,d} \to \mu^+\mu^-$
processes can be also modified from the SM predictions.
If the second generation masses are modified, the $K$-$\bar K$ and/or $D$-$\bar D$ mixings 
are modified.
One may adjust them to the flavor data in general even if there are large corrections in all the quark masses and mixings,
but it is not an attractive situation. 
Among the masses and mixings, only the first generation masses can be modified
without a major contradiction to the flavor data,
and thus, it is suitable to make the minimal modification to the fit results.

The finite correction of the down-type quark masses from the gluino loop diagram 
(neglecting the effect from flavor mixings) can be written as
\begin{equation}
\Delta m_{d_i} \simeq  \frac{2\alpha_s}{3\pi} m_{\tilde g} \sin\theta_{LR}^i \left(F\left(\frac{m^2_{\tilde g}}{m^2_{\tilde d_L^i}}\right) -F\left(\frac{m^2_{\tilde g}}{m^2_{\tilde d_R^i}}\right)\right),
\end{equation}
where $m_{\tilde g}$ is a gluino mass, $F(x) = \log x/(x-1)$,
and $\theta_{LR}$ is a left-right mixing angle,
\begin{equation}
\sin\theta_{LR}^i \simeq \frac{A_d^i v_d - \mu m_{d_i} \tan\beta}{m^2_{\tilde d_L^i} - m^2_{\tilde d_R^i}}.
\end{equation}
Thus,
if the $A$ term is proportional to the Yukawa coupling as in the minimal supergravity model ($A_d^i v_d \simeq A m_{d_i}$),
the gluino contribution to the fermion mass is flavor universal :
\begin{equation}
\left(\frac{\Delta m_d}{m_d}\right)_{\rm gluino} 
\simeq \left(\frac{\Delta m_s}{m_s}\right)_{\rm gluino}  \simeq \left(\frac{\Delta m_b}{m_b}\right)_{\rm gluino} .
\end{equation}
The chargino contribution can break the flavor universality to the mass corrections,
but the corrections to down and strange quark are still universal.
In order to correct the fit of the down quark masses and mixings,
we need to break the universality of the corrections, and therefore, the non-proportional term
of the trilinear coupling is needed.

In the minimal supergravity model,
the scalar potential contains the trilinear coupling originated from the mixed term between the visible
and hidden sectors, $W= W_{\rm vis} + W_{\rm hid}$, 
and thus, the scalar trilinear couplings are proportional to the Yukawa couplings.
In the string-inspired models, the Yukawa coupling can depends on moduli, $\tau$,
and the non-proportional term can be induced\footnote{
In general, denoting the cubic coupling of the superfield $\Phi_I$ in the superpotential as $Y_{IJK} \Phi_I \Phi_j \Phi_K$,
one obtains the scalar trilinear coupling as
\begin{equation}
A_{IJK} = A_0 (Y_{IJK} + c\, \partial_\tau Y_{IJK}(\tau)).
\end{equation}
}.
The Yukawa matrix and the derivative of the Yukawa matrix are not necessarily simultaneously diagonalized,
and therefore, the non-proportional term can have complete different structure compared to the fermion mass matrix.
As a result, the (1,1) element of the non-proportional part of the $A$ term (in the basis of the Yukawa matrix is diagonal)
 is not necessarily hierarchically small, and it can easily modify the first generation fermion masses.
 If the gluino and squark masses are around 2 TeV,
we estimate that $A_d^{11} \cos\beta$ is sub-GeV to modify the down quark mass.
Suppose that the all the components are roughly same order of the trilinear coupling,
the corrections to the other elements of the mass matrix are also of the order of 1 MeV, which can be negligible
for the generation mixings and the FCNC processes.

If the down quark mass is modified via the trilinear scalar coupling,
the up quark mass can be also modified.
Therefore, the (1,1) element of the trilinear coupling for the up-type quarks should be small,
which is possible within a freedom of the model parameter.
Because there are only two Higgs coupling matrices in the minimal model,
the (1,1) element of the trilinear coupling for the charged lepton remains in that case,
and thus, the electron mass is also modified. 
The correction to the electron mass is induced by the Bino component of the 
neutralino, and the modification is not large.
Instead, there can have freedom to tune the loop correction of the up quark mass and 
cancel the tree-level one to make the physical up quark mass tiny.

The scalar trilinear triplet Higgs couplings can induce
dimension-four proton decay operators via the triplet Higgs scalar exchange,
and the (1,1) element of the coupling can make the excess of the proton decay amplitudes
after double gaugino dressing.
However, to generate the dimension-four operators, 
the holomorphic bilinear mass term of the triplet Higgs scalars is needed.
Besides, the other elements of the trilinear scalar coupling matrix are free and can be small.
Thus, the induced amplitudes can be suppressed within the freedom of the parameters.

As the implication to the low energy observables,
the bounds from the neutron EDM need to be considered,
as in the case of the minimal flavor violation,
and the non-proportional term of the trilinear coupling should be hermitian
(in the basis where the quark mass matrix is real/positive diagonal).
Because the coupling matrix is symmetric due to the $SO(10)$ algebra,
the phases of the components in the non-proportional term are severely restricted.
The non-proportional term can be a new source of the flavor violation 
via the RGE evolution of the scalar masses,
and the $\mu \to e\gamma$ process will the most stringent constraint on it.
As described before, the non-proportional term is not necessarily have a hierarchical
structure parallel to the Yukawa hierarchy, and within the freedom,
the contributions to $\mu\to e\gamma$ can be switched off.
Rather, choosing the size of (1,2)/(2,1) elements of the trilinear coupling,
the size of the $\mu\to e\gamma$ amplitude can be adjusted,
and the production studied in the previous sections becomes just a reference value.
However, the qualitative statement relating to the $v_R$ scale dependence 
is expected to be kept.

We note that 
the charge/color breaking (CCB) vacua may appear for the large (1,1) element of the scalar trilinear coupling,
and the electroweak symmetry breaking vacua may become metastable \cite{Casas:1996de,Park:2010wf},
similarly to the models in which the fermion masses and mixings are modified by the loop corrections \cite{Iskrzynski:2014sba}.
Surely, if the lifetime of our vacuum can be larger than the age of the universe \cite{Kusenko:1996jn}, the model is still valid.
The bounds from the tunneling processes of the vacua should be analyzed model-independently,
and the bounds are beyond the scope of this paper. We just mention that the vacua may be unstable,
but long-lived as in the other analyses.

\section{Conclusions and discussions}
We study the minimal $SO(10)$ model in which quarks and leptons couple to only $\bf 10$ and $\overline{\bf 126}$ Higgs representations. 
In the minimal Yukawa coupling, the up quark mass is not necessarily small, and as a consequence, the relation $m_e m_\mu m_\tau \simeq m_d m_s m_b$ is not obvious for a proper size of Cabibbo angle \cite{Dutta:2009ij}. 
However, the quark and lepton masses can be fully fitted (without such a clear reason) by using the full parameter freedom even in the case of the minimal Yukawa coupling. 
In fact, as we have described, in the case of $R\sim O(1)$ (i.e. $v_R \sim 10^{13}$ GeV), the freedom is active to fit all of the fermion masses and mixings, which is consistent with the number counting of the parameters.

Even though the fitting of fermion masses and mixings is just a choice of parameters,
the minimal $SO(10)$ is still attractive since the model is predictive. As was seen, the scale of $v_R$ is important for the $\chi^2$ minimum fits, and it is related to an intermediate scale determined by the $SO(10)$ symmetry breaking vacua.
Although we perform the search of the $\chi^2$ minimum 
by using the fermion masses at the GUT scale in the minimal SUSY boundary conditions,
the qualitative properties of the solutions described in Section 3 are also applied to the 
case of non-SUSY boundary conditions.
In non-SUSY model, the intermediate scale  is rather favored
for the gauge coupling unification, contrary to the SUSY models,
and 
the constraints from the dimension-five proton decays
and the flavor physics are of course absent.
To say it inversely,
the predicted fermion mass matrices cannot be probed nor applied to the low energy physics
in the non-SUSY model.
In the sense to probe the GUT scale physics, 
the SUSY scenario of the predictive minimal Yukawa coupling in the $SO(10)$ model
is still attractive and worth to be chased.
As it was explained,
the solution for $v_R \sim 10^{16}$ GeV is interesting to apply
even though the threshold corrections are needed to fit the center value of the down quark mass.

As methodology, physicists often assume the minimality of a model,
not only because the model is predictive due to the reduced number of the parameters,
but also because the prototypal model can contain the essential features to study the phenomenology of the objects.
In general, it is obvious that the predictivity can be obtained by reducing the number of parameters.
It is important to dissect the property of the objects to see whether it only originates from the minimality
or whether it comes from any other fundamental mechanism theoretically.
In reality, the minimality at the GUT scale physics is hard to be verified experimentally,
and one can always claim that the corrections can be induced since the reduced Planck scale is just two digit above 
the GUT scale.
Therefore, the concern one has to care about is whether the prediction is stable under the perturbation of the minimality assumption
in GUT models,
and we should investigate both the speciality of the prediction from the minimality assumption  
and the essential features which remain predictable even after the model is extended.
As described, the major prediction of the minimal $SO(10)$ model (in which the Yukawa coupling to the fermions is minimal) is that
the Dirac and Majorana neutrino mass matrices are fully determined.

Sometimes, the Higgs contents are also restricted to be minimal, i.e. ${\bf 10}+{\bf 126}+ \overline{\bf 126}+ {\bf 210}$,
and the number of parameters in the Higgs superpotential is made to be minimal in the $SO(10)$ model.
In the SUSY version of the minimal Higgs contents,
the fermion data fittings require a typical $SO(10)$ symmetry breaking vacuum, 
 and some decomposed particles from the Higgs multiplets appear at the intermediate scale to obtain the proper size of neutrino masses.
The gauge coupling evolution is severely affected by the presence of the particles at the intermediate scales,
and the gauge unification suggests a non-SUSY (or split-SUSY) version of the model \cite{Bertolini:2006pe,Bajc:2008dc,Bertolini:2009es}.
In non-SUSY scenario, however,
it is impossible to communicate with the predicted fermion mass matrices
by using the low energy phenomena, such as the various modes of baryon number violating processes and lepton flavor violation,
which are induced by the Yukawa interactions.
We, thus, take a stance to probe the footprint of the GUT scale physics by 
using the SUSY scenario, in which the whole picture of the predicted mass matrices can be influenced to the low energy physics.
In order to have the stance, 
we do not employ the minimality of the Higgs contents.
As we have described frequently in this paper,
we define the minimal $SO(10)$ model as that based on the minimality of the Yukawa interaction,
which is fully predictive to find the structure of the fermion mass matrices.

The important parameters in SUSY GUTs, which can affect low energy physics,
 are ${\bf 126}$ vev $v_R$, $SU(2)_L$ triplet Higgs mass $M_\Delta$, the lightest colored Higgs mass $M_{H_C}$, and the heavy gauge boson mass $M_G$.
 The parameters, $v_R$, $M_\Delta$, $M_{H_C}$, and $M_G$, are functions of the parameters 
 in the Higgs potential and particle spectrum (depending on the consistency of the gauge coupling evolutions) deductively,
 and one can discuss how the parameters can depend on $SO(10)$ breaking vacua and the particle spectrum \cite{Dutta:2007ai}.
 Instead of specifying the Higgs potential and deriving the scale $v_R$ 
from the certain number of parameters in the Higgs potential, we choose them to be free parameters as methodology
because there are plenty of choices to extend the Higgs potential.
In fact, $v_R$ is an important scale to describe the GUT symmetry breaking and the neutrino mass spectrum,
and it is an open question how the right-handed Majorana neutrino is generated in GUTs
and the size of the light neutrino masses are obtained.
As we have seen, the fit of the fermion masses and mixings depends on $v_R$.
Actually, 
the naive scale of the right-handed neutrino masses to reproduce the observation
is less than $10^{14}$ GeV, which is hierarchically smaller than the GUT scale.
In other words, the neutrino masses are a bit heavier if the GUT scale, $10^{16}$ GeV, is 
considered to be a fundamental scale,
and explaining the hierarchy is one of the open questions in GUTs.
In the solution of the minimal $SO(10)$ model featured in this 
paper, the nearly singular Majorana mass matrix is contained for a 
large scale of $v_R$ instead of the smallness of the each component of the matrix.
The nearly singular Majorana matrix is one of the solution to explain
the neutrino masses keeping the symmetry breaking scale to be $10^{16}$ GeV,
because the inverse of the Majorana mass matrix is included the type I seesaw formulae. 
In that sense, this solution can provide an approach to the fundamental question,
and it is interesting to research the prediction of the model.

In $SU(5)$ GUT, the Dirac neutrino mass matrix is completely independent to the other charged fermions.
In the unified model scenario, the hierarchical forms of the Dirac and Majorana neutrino mass matrices are often assumed.
The Dirac neutrino mass is generated by the couplings to the up-type Higgs,
and in a certain model, the hierarchy of the Dirac neutrino mass is assumed to be similar to the up-type quarks. 
On the other hand, the predicted Dirac neutrino Yukawa coupling in the minimal $SO(10)$ model is less hierarchical rather 
than the up-type quarks, and the hierarchy 
rather resembles to the down-type quark or charged-leptons.
This is simply because the up-type quark mass hierarchy in this model
is realized by a cancellation between two Yukawa matrices.
Even in the non-minimal models, in which $\bf 120$ Higgs is added to the Yukawa couplings 
and the (1,1) elements of the Yukawa couplings are assumed to be small to suppress proton decay amplitudes in an $SO(10)$ model,
the Dirac neutrino Yukawa coupling also resembles to the down-type quarks and the charged-leptons 
and the size of the induced FCNCs is similar to the minimal model \cite{Dutta:2013bvf}.
In that sense, the size of the induced FCNCs is predictable in the renormalizable $SO(10)$ models.
This feature provides an important implication to the process of $\mu\to e\gamma$ in SUSY models,
and it can be tested by the experiment.

As described in the text, the electron EDM is one of the major implications of the $v_R$ dependence in the $SO(10)$ model.
If the scale $v_R$ is less than the unification scale, 
the flavor violation is generated in both left- and right-handed slepton mass matrices,
and the electron EDM is enlarged.
If the source of the flavor violation is hermitian, the phase of the EDM amplitude is canceled. 
In the minimal model, the Yukawa coupling is symmetric due to the $SO(10)$ algebra,
and the components are not real numbers to reproduce the KM phase.
Therefore, the electron EDM can be induced in general for the lower scale $v_R$.
On the other hand, the Yukawa coupling to the $\bf 120$ representation is anti-symmetric,
and thus, there can be room to generate the hermitian flavor violation in the non-minimal models.
Consequently, the electron EDM can be a major probe to distinguish the models.

Among the quark-lepton phenomena directly induced by the Yukawa interaction,
the CP violation in the neutrino oscillations is the last piece to be discovered.
In fact, the ongoing experiments have already given the hint that 
 CP is violated in the oscillations,
and it is expected that the CP phase (PMNS phase) is measured in the near future.
Remarkably, the minimal $SO(10)$ model has predicted the proper size of neutrino 13-mixing angle (before its measurement),
which is consistent with the current experimental results.
We show the prediction of the PMNS phase in the minimal $SO(10)$ model,
in which there are disfavored region depending on the 23-mixing.
Now the prediction of the PMNS phase of the model is being challenged to the
experimental results, including the precise measurement of the 23-mixing.

\appendix

\section{Appendix : Square root matrix}

A square root matrix of $A$, denoted as $\sqrt{A}$, is defined to be matrices which satisfy
$\left(\sqrt{A}\right)^2 =A$.

In our numerical practical reason, we assume that the eigenvalues of a $3\times 3$ matrix $A$, $\lambda_{1,2,3}$,
 are not degenerate.
Then, it can be diagonalized by a matrix $V$:
\begin{equation}
A = V 
\left(
  \begin{array}{ccc}
   \lambda_1 & & \\
   & \lambda_2 & \\
   & & \lambda_3 \\
  \end{array}
 \right)
 V^{-1}.
\end{equation}
The square root matrix can be expressed as
\begin{equation}
\sqrt{A} = V 
\left(
  \begin{array}{ccc}
   \pm \sqrt{\lambda_1} & & \\
   & \pm \sqrt{\lambda_2} & \\
   & & \pm \sqrt{\lambda_3} \\
  \end{array}
 \right)
V^{-1}.
\end{equation}
The double-signs are in no particular order (INPO),
and there are 8-fold matrices for the square root of a $3\times 3$ matrix.

Defining matrices $\Lambda_i$ as
\begin{equation}
\Lambda_1 = V 
 \left(
  \begin{array}{ccc}
   1 & & \\
   & 0 & \\
   & & 0 \\
  \end{array}
 \right)
 V^{-1}, \quad
\Lambda_2 = V \left(
  \begin{array}{ccc}
   0 & & \\
   & 1 & \\
   & & 0 \\
  \end{array}
  \right)
 V^{-1}, \quad
\Lambda_3 = V \left( 
  \begin{array}{ccc}
   0 & & \\
   & 0 & \\
   & & 1 \\
  \end{array}
\right)
V^{-1},
\end{equation}
we obtain
\begin{equation}
A = \lambda_1 \Lambda_1 + \lambda_2 \Lambda_2 + \lambda_3 \Lambda_3.
\end{equation}
One can easily obtain the following relations:
\begin{equation}
\Lambda_1 + \Lambda_2 + \Lambda_3 = {\bf 1}, \qquad \Lambda_i \Lambda_j = \Lambda_{i} \delta_{ij}.
\end{equation}
By solving the simultaneous equation
\begin{equation}
\left(
 \begin{array}{ccc}
  1 & 1 & 1 \\
  \lambda_1 & \lambda_2 & \lambda_3 \\
  \lambda_1^2 & \lambda_2^2 & \lambda_3^2 
 \end{array}
\right)
\left(
\begin{array}{c}
 \Lambda_1 \\
 \Lambda_2 \\
 \Lambda_3
\end{array}
\right)
=
\left(
\begin{array}{c}
 {\bf1} \\
 A \\
 A^2
\end{array}
\right),
\end{equation}
we obtain
\begin{eqnarray}
\Lambda_1 &=& \frac{1}{(\lambda_1 - \lambda_2)(\lambda_1 - \lambda_3)}
\left(
 A^2 - (\lambda_2 + \lambda_3) A + \lambda_2 \lambda_3 {\bf1}
\right), \\
\Lambda_2 &=& \frac{1}{(\lambda_2 - \lambda_1)(\lambda_2 - \lambda_3)}
\left(
 A^2 - (\lambda_1 + \lambda_3) A + \lambda_1 \lambda_3 {\bf1}
\right), \\
\Lambda_3 &=& \frac{1}{(\lambda_3 - \lambda_1)(\lambda_3 - \lambda_2)}
\left(
 A^2 - (\lambda_1 + \lambda_2) A + \lambda_1 \lambda_2 {\bf1}
\right).
\end{eqnarray}

Using the matrices $\Lambda_i$, 
the square root matrix can be written as
\begin{equation}
\sqrt{A} = \sum_i s_i \sqrt{\lambda_i} \Lambda_i,
\end{equation}
where $s_i$ stands for the sign of each root of eigenvalue,
$s_i = \pm 1$.


\section*{Acknowledgments}

The work of T.F.\ is supported in part by the Grant-in-Aid for
Science Research from the Ministry of Education, Science and
Culture of Japan (No.\ 26247036).
The work of K.I. is supported in part by JSPS Research Fellowships for Young Scientists. 
The work of Y.M. is supported by the Excellent Research Projects of
 National Taiwan University under grant number NTU-EPR-104R8915. 


\end{document}